\documentclass[12pt]{article}

\usepackage{graphicx}

\def\Fint{\rlap{$\Biggl\rfloor$}\Biggl\lceil}

\def\square{\kern1pt\vbox{\hrule height 1.2pt\hbox{\vrule width 1.2pt\hskip 3pt
   \vbox{\vskip 6pt}\hskip 3pt\vrule width 0.6pt}\hrule height 0.6pt}\kern1pt}

\begin{document}

\begin{titlepage}

\begin{flushright}
UFIFT-QG-12-02
\end{flushright}

\vspace{1cm}

\begin{center}
{\bf Graviton Corrections to Maxwell's Equations}
\end{center}

\vspace{.5cm}

\begin{center}
Katie E. Leonard$^{\dagger}$ and R. P. Woodard$^{\ddagger}$
\end{center}

\vspace{.5cm}

\begin{center}
\it{Department of Physics \\
University of Florida \\
Gainesville, FL 32611}
\end{center}

\vspace{1cm}

\begin{center}
ABSTRACT
\end{center}
We use dimensional regularization to compute the one loop quantum
gravitational contribution to the vacuum polarization on flat space
background. Adding the appropriate BPHZ counterterm gives a fully
renormalized result which we employ to quantum correct Maxwell's
equations. These equations are solved to show that dynamical photons
are unchanged, provided the free state wave functional is
appropriately corrected. The response to the instantaneous
appearance of a point dipole reveals a perturbative version of the
long-conjectured, ``smearing of the light-cone''. There is no change
in the far radiation field produced by an alternating dipole.
However, the correction to the static electric field of a point
charge shows strengthening at short distances, in contrast to 
expectations based on the renormalization group. We check for gauge
dependence by working out the vacuum polarization in a general
3-parameter family of covariant gauges.

\vspace{.5cm}

\begin{flushleft}
\end{flushleft}

\vspace{1cm}

\begin{flushleft}
$^{\dagger}$ e-mail: katie@phys.ufl.edu \\
$^{\ddagger}$ e-mail: woodard@phys.ufl.edu
\end{flushleft}
\end{titlepage}

\section{Introduction}

Electromagnetism provided the first example of a relativistic,
unified gauge theory. Later on, it was quantum electrodynamics (QED)
which produced the first quantitative successes in the struggle to
understand interacting quantum field theories. It is therefore
natural to wonder what electromagnetism can tell us about quantum
gravity.

\begin{figure}[b]
\begin{center}
\includegraphics[width=5.0cm,height=2.5cm]{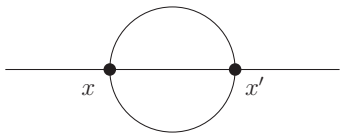}
\end{center}
\caption{A two loop contribution to the self-mass-squared in $\lambda
\phi^4$ theory.}
\label{setsun}
\end{figure}

Efforts along these lines date back more than half a century, and
were at first concerned with a phenomenon termed, ``smearing of the
light-cone'' \cite{smear}. The idea is that quantum gravitational
effects might soften the divergences of other quantum field theories
because those divergences are associated with the singularities all
propagators develop for null separations \cite{DWB},
\begin{equation}
i\Delta[g](x;x') = \frac1{2 \pi^2} \frac1{\sigma[g](x;x')} +
\mathcal{O}\Bigl( \ln(\sigma)\Bigr) \; .
\end{equation}
Here $i\Delta[g](x;x')$ is the scalar propagator in the presence of a
general metric background $g_{\mu\nu}$, and $\sigma[g](x;x')$ is
$\frac12$ times the square of the geodesic length from $x^{\mu}$ to
${x'}^{\mu}$ in that metric. Although the propagator is a well-defined
distribution --- its 4-dimensional integral against a test function
converges --- powers of it are not. That is why there is a quadratic
ultraviolet divergence in the two loop ``setting sun'' contribution to
the $\lambda \phi^4$ self-mass-squared depicted in Fig.~\ref{setsun},
\begin{equation}
-iM^2_{\rm s.s}(x;x') = \frac{(-i\lambda)^2}{3!} \, \sqrt{-g(x)} \,
\Bigl[ i\Delta[g](x;x') \Bigr]^3 \sqrt{-g(x')} \; . \label{GRssun}
\end{equation}
For fixed $x^{\mu}$, the singularity occurs at different points ${x'}^{\mu}$
as the metric $g_{\mu\nu}$ is varied. Quantizing gravity entails
functionally averaging (\ref{GRssun}) over metrics, and this might be
expected to reduce or eliminate the singularity.

\begin{figure}[b]
\begin{center}
\includegraphics[width=5.0cm,height=1.25cm]{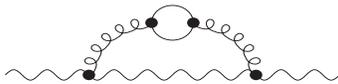}
\end{center}
\caption{Lowest order gravitational effect on light due to quantum
fluctuations of matter. Photon lines are wavy, graviton lines are
winding, and matter lines are solid.}
\label{stress}
\end{figure}

Of course smearing the light-cone would distort the propagation of
light, and even a tiny angular deviation might show up over
cosmological distances. Metric fluctuations can also induce
luminosity and redshift variations, as well as spectral line
broadening and angular blurring. These effects have typically been
studied indirectly, by working out the scattering of light by a
classical metric \cite{Zipoy}. A significant example is the
computation Sachs and Wolfe made of the anisotropies cosmic
microwave photons acquire in propagating through primordial
cosmological perturbations \cite{Sachs}. More recent treatments
have been given of the effects of various ensembles of gravitons
\cite{Beilok}. (It is but a short step to adjust the distribution
to reproduce quantum 0-point fluctuations, but this does not seem
to have been done.) And there has been much recent interest in
metric fluctuations engendered by quantum oscillations of matter
fields \cite{Ford}, a representative diagram for which is given
in Fig.~\ref{stress}.

Quantum gravity also affects electrodynamic forces. Radkowski
seems to have performed the first computation of the one loop
correction to the Coulomb potential of a point charge \cite{AFR}.
A somewhat different result was inferred from the scattering
of charged scalars \cite{Coulomb}. Both of those studies found
that quantum gravity strengthens the electrostatic force at short
distances. Much recent interest attended the proposal of Robinson
and Wilczek that the effect goes the other way, based on
renormalization group flows \cite{RW}.

A widely perceived impediment to all of these studies is the fact
that general relativity plus electromagnetism is not perturbatively
renormalizable, even at one loop order \cite{SDVN}. One consequence
is that we cannot compute everything reliably, but that does not
mean we are unable to calculate anything. One must simply view
quantum gravity as a low energy effective field theory whose
divergences are absorbed using BPHZ (Bogoliubov, Parsiuk,
Hepp and Zimmermann) counterterms \cite{BPHZ}. In this context,
loops of massless gravitons and photons engender nonlocal and
ultraviolet finite contributions to the effective action which
are unique predictions of the theory that cannot be changed by its
ultraviolet completion \cite{JFD}. (For example, see the string theory result \cite{EKCK}.) This is why Bloch and Nordsieck
were able to resolve the infrared problem of QED \cite{BN}, long
before that theory's renormalizablility was understood. It is also
why Weinberg was able to derive a similar resolution for the infrared
problem of quantum gravity \cite{SW}, and why Feinberg and Sucher
were able to use Fermi theory to compute the long range force
engendered by the exchange of massless neutrinos \cite{FS}.

One is of course free to criticize the study of quantum gravity
plus electromagnetism on the grounds that the predicted effects
are too small to be observable. Our own interest in the subject
derives from its relevance as the flat space correspondence limit
of the regime of primordial inflation, during which quantum
gravitational effects are not unobservably small. Indeed, the
gravitational response to quantum fluctuations of matter \cite{Slava}
has been resolved \cite{WMAP}, and the corresponding fluctuations of
gravitational radiation \cite{AAS} may soon be detected \cite{PLANCK}.

Whether during primordial inflation or on flat space background, the
proper vehicle for studying quantum distortions of electrodynamics
is the quantum-corrected Maxwell equation. One gets this by first
computing the ``vacuum polarization'' $i[\mbox{}^{\mu}\Pi^{\nu}](x;x')$,
which is the one-particle-irreducible (1PI) 2-point function for the
photon. This is then used to quantum-correct Maxwell's equation,
\begin{equation}
\partial_{\nu} \Bigl[ \sqrt{-g} \, g^{\nu\rho} g^{\mu\sigma}
F_{\rho\sigma}(x) \Bigr] + \int \!\! d^4x' \, \Bigl[\mbox{}^{\mu}
\Pi^{\nu}\Bigr](x;x') A_{\nu}(x') = J^{\mu}(x) \; . \label{qmax1}
\end{equation}
This framework has been employed to infer the effects of inflationary
charged scalar production on photons \cite{SQED}, and on electrodynamic
forces \cite{Henri}. The purpose of this paper is to facilitate a
similar study of the effects of inflationary graviton production by
first working out the flat space correspondence limit. An example of
the utility of this exercise is the recent examination of the effects
of inflationary scalars on gravitons \cite{PW1}, for which the flat
space limit \cite{PW2} provided crucial guidance in dealing with the
vastly more complicated graviton self-energy \cite{PW3} that pertains
during primordial inflation.

The quantum gravitational contribution to the one loop vacuum
polarization is derived in section 2. Section 3 solves the
quantum-corrected Maxwell equation (\ref{qmax1}) for photons, for the
instantaneous creation of a point dipole, for an alternating point
dipole, and for a static point charge. The issue of gauge dependence
is discussed in section 4, and our conclusions comprise section 5.

\section{One Loop Vacuum Polarization}

The purpose of this section is to compute the renormalized, one loop
contribution to the vacuum polarization from quantum gravity on flat
space background. We begin by presenting the necessary Feynman rules.
Then we use them to compute the dimensionally regulated result. By a
process of successive partial integrations this is expressed as a
divergent, local term --- which is canceled by a BPHZ counterterm ---
plus the finite, nonlocal contribution which constitutes the
renormalized result.

\subsection{Feynman Rules}

Our total Lagrangian contains three parts,
\begin{equation}
\mathcal{L} = \mathcal{L}_{\rm GR} + \mathcal{L}_{\rm EM} +
\mathcal{L}_{\rm BPHZ} \; .
\end{equation}
These are, respectively, the Lagrangians of general relativity,
electromagnetism and the BPHZ counterterm required for this
computation,
\begin{eqnarray}
\mathcal{L}_{\rm GR} & = & \frac1{16 \pi G} \, R \sqrt{-g} \; , 
\label{LGR} \\
\mathcal{L}_{\rm EM} & = & -\frac14 F_{\mu\nu} F_{\rho \sigma}
g^{\mu\rho} g^{\nu \sigma} \sqrt{-g} \; , \label{LEM} \\
\mathcal{L}_{\rm BPHZ} & = & C_4 D_{\alpha} F_{\mu\nu} D_{\beta}
F_{\rho\sigma} g^{\alpha\beta} g^{\mu\rho} g^{\nu\sigma} \sqrt{-g} \; .
\label{LBPHZ}
\end{eqnarray}
We employ a $D$-dimensional, spacelike metric $g_{\mu\nu}$, with
inverse $g^{\mu\nu}$ and determinant $g = {\rm det}(g_{\mu\nu})$.
Our affine connection and Riemann tensor are,
\begin{eqnarray}
\Gamma^{\rho}_{~\mu\nu} & \equiv & \frac12 g^{\rho\sigma} \Bigl[
\partial_{\nu} g_{\sigma\mu} \!+\! \partial_{\mu} g_{\nu\sigma} \!-\!
\partial_{\sigma} g_{\mu\nu} \Bigr] \; , \\
R^{\rho}_{~\sigma\mu\nu} & \equiv & \partial_{\mu} \Gamma^{\rho}_{~\nu\sigma}
\!-\! \partial_{\nu} \Gamma^{\rho}_{~\mu\sigma} \!+\!
\Gamma^{\rho}_{~\mu\alpha} \Gamma^{\alpha}_{~\nu\sigma} \!-\!
\Gamma^{\rho}_{~\nu\alpha} \Gamma^{\alpha}_{~\mu\sigma} \; . \qquad
\end{eqnarray}
Our Ricci tensor is $R_{\mu\nu} \equiv R^{\rho}_{~\mu\rho\nu}$ and the
associated Ricci scalar is $R \equiv g^{\mu\nu} R_{\mu\nu}$. The
electromagnetic field strength tensor and its first covariant derivative
are,
\begin{eqnarray}
F_{\mu\nu} & \equiv & \partial_{\mu} A_{\nu} \!-\! \partial_{\nu}
A_{\mu} \; , \\
D_{\alpha} F_{\mu\nu} & \equiv & \partial_{\alpha} F_{\mu\nu} \!-\!
\Gamma^{\gamma}_{~\alpha\mu} F_{\gamma\nu} \!-\!
\Gamma^{\gamma}_{~\alpha\nu} F_{\mu\gamma} \; . \qquad
\end{eqnarray}

We define the graviton field $h_{\mu\nu}(x)$ as the difference
between the full metric and its Minkowski background value
$\eta_{\mu\nu}$,
\begin{equation}
g_{\mu\nu}(x) \equiv \eta_{\mu\nu} + \kappa h_{\mu\nu}(x) \; ,
\end{equation}
where $\kappa^2 \equiv 16 \pi G$ is the loop counting parameter of
quantum gravity. We follow the usual conventions whereby a comma
denotes ordinary differentiation, the trace of the graviton field is
$h \equiv \eta^{\mu\nu} h_{\mu\nu}$, and graviton indices are raised
and lowered using the Minkowski metric, $h^{\mu}_{~\nu} \equiv
\eta^{\mu\rho} h_{\rho\nu}$ and $h^{\mu\nu} \equiv \eta^{\mu\rho}
\eta^{\nu\sigma} h_{\rho \sigma}$. After extracting a surface term
the gravitational Lagrangian can be written as,
\begin{eqnarray}
\lefteqn{\mathcal{L}_{\rm GR} - {\rm Surface} = \sqrt{-g} \,
g^{\alpha\beta} g^{\rho\sigma} g^{\mu\nu} } \nonumber \\
& & \hspace{2cm} \times \Biggl\{ \frac12 h_{\alpha\rho ,\mu}
h_{\nu\sigma ,\beta} \!-\! \frac12 h_{\alpha\beta ,\rho}
h_{\sigma\mu ,\nu} \!+\! \frac14 h_{\alpha\beta ,\rho}
h_{\mu\nu ,\sigma} \!-\! \frac14 h_{\alpha\rho ,\mu}
h_{\beta\sigma ,\nu} \Biggr\} . \qquad \label{Linv}
\end{eqnarray}

The quadratic part of the invariant Lagrangian is,
\begin{equation}
\mathcal{L}^{(2)}_{\rm GR} = \frac12 h^{\rho\sigma , \mu}
h_{\mu \sigma , \rho} \!-\! \frac12 h^{\mu \nu}_{~~ ,\mu} h_{,\nu}
\!+\! \frac14 h^{,\mu} h_{, \mu} \!-\! \frac14 h^{\rho\sigma , \mu}
h_{\rho \sigma , \mu} \; .
\end{equation}
We fix the gauge by adding,
\begin{equation}
\mathcal{L}_{\rm GRfix} = -\frac12 \eta^{\mu\nu} F_{\mu} F_{\nu} \qquad ,
\qquad F_{\mu} \equiv \eta^{\rho\sigma} \Bigl(h_{\mu\rho , \sigma}
- \frac12 h_{\rho \sigma , \mu} \Bigr) . \label{GRfix}
\end{equation}
The resulting graviton propagator can be expressed in terms of the
massless scalar propagator $i\Delta(x;x')$,
\begin{equation}
i \Bigl[\mbox{}_{\mu\nu} \Delta_{\rho\sigma}\Bigr](x;x') =
\Bigl[ 2 \eta_{\mu (\rho} \eta_{\sigma) \nu} \!-\! \frac2{D \!-\! 2}
\eta_{\mu\nu} \eta_{\rho\sigma} \Bigr] i\Delta(x;x') \; . \label{Gravprop}
\end{equation}
The spacetime dependence of the scalar propagator derives from the
Lorentz interval $\Delta x^2(x;x')$,
\begin{equation}
\Delta x^2(x;x') \equiv \Bigl\Vert \vec{x} \!-\! \vec{x}'
\Bigr\Vert^2 \!-\!  \Bigl( \vert t \!-\! t'\vert \!-\! i \varepsilon
\Bigr)^2 \quad \Longrightarrow \quad i\Delta(x;x') =
\frac{\Gamma(\frac{D}2 \!-\! 1)}{4 \pi^{\frac{D}2} } \Bigl(
\frac1{\Delta x^2} \Bigr)^{\frac{D}2 - 1} \; . \label{scalprop}
\end{equation}

The quadratic part of the electromagnetic action is,
\begin{equation}
\mathcal{L}_{\rm EM} = -\frac12 \partial_{\mu} A_{\nu} \partial^{\mu}
A^{\nu} + \frac12 (\partial_{\mu} A^{\mu})^2 \; .
\end{equation}
We fix the gauge by adding,
\begin{equation}
\mathcal{L}_{\rm EMfix} = -\frac12 (\partial_{\mu} A^{\mu})^2 \; .
\label{EMfix}
\end{equation}
The associated photon propagator is,
\begin{equation}
i \Bigl[ \mbox{}_{\mu} \Delta_{\rho}\Bigr](x;x') = \eta_{\mu\rho} \,
i\Delta(x;x') \; . \label{Photprop}
\end{equation}

Electromagnetic interaction vertices descend from the second variational
derivative of the action,
\begin{equation}
\frac{\delta^2 S_{\rm EM}}{\delta A_{\mu}(x) \delta A_{\rho}(x') }
= \partial_{\kappa} \Biggl\{ \sqrt{-g(x)} \, \Bigl[ g^{\kappa\lambda}(x)
g^{\mu \rho}(x) \!-\! g^{\kappa\rho}(x) g^{\lambda\mu}(x) \Bigr]
\partial_{\lambda} \delta^D(x \!-\! x') \Biggr\} \; .
\end{equation}
The necessary vertex functions are obtained by expanding the metric
factors,
\begin{eqnarray}
\lefteqn{\sqrt{-g} \, \Bigl( g^{\kappa\lambda} g^{\mu\rho} \!-\!
g^{\kappa\rho} g^{\lambda\mu} \Bigr) \equiv \eta^{\kappa\lambda}
\eta^{\mu\rho} \!-\! \eta^{\kappa\rho} \eta^{\lambda\mu} } \nonumber \\
& & \hspace{3.5cm} + \kappa V^{\mu\rho\kappa\lambda\alpha\beta}
h_{\alpha\beta} + \kappa^2 U^{\mu\rho\kappa\lambda\alpha\beta\gamma\delta}
h_{\alpha\beta} h_{\gamma\delta} + O(\kappa^3) \; . \qquad
\end{eqnarray}
The 3-point and 4-point vertices are,
\begin{eqnarray}
\lefteqn{V^{\mu\rho\kappa\lambda\alpha\beta} = \eta^{\alpha\beta}
\eta^{\kappa [\lambda} \eta^{\rho ] \mu} \!+\! 4 \eta^{\alpha) [\mu}
\eta^{\kappa ] [\rho} \eta^{\lambda ] (\beta} \; , } \label{Vvert} \\
\lefteqn{U^{\mu\rho\kappa\lambda\alpha\beta\gamma\delta} = \Bigl[\frac14
\eta^{\alpha\beta} \eta^{\gamma\delta} \!-\! \frac12 \eta^{\alpha (\gamma}
\eta^{\delta) \beta} \Bigr] \eta^{\kappa [\lambda} \eta^{\rho ] \mu} +
\eta^{\alpha\beta} \eta^{\gamma) [\mu} \eta^{\kappa] [\rho}
\eta^{\lambda] (\delta} } \nonumber \\
& & \hspace{-.5cm} + \eta^{\gamma\delta} \eta^{\alpha) [\mu} \eta^{\kappa]
[\rho} \eta^{\lambda] (\beta} \!+\! \eta^{\kappa (\alpha} \eta^{\beta)
[\lambda} \eta^{\rho ] (\gamma} \eta^{\delta) \mu} \!+\! \eta^{\kappa (\gamma}
\eta^{\delta) [\lambda} \eta^{\rho ] (\alpha} \eta^{\beta) \mu}
\!+\! \eta^{\kappa (\alpha} \eta^{\beta) (\gamma} \eta^{\delta) [\lambda}
\eta^{\rho ] \mu} \nonumber \\
& & \hspace{1.5cm} + \eta^{\kappa (\gamma} \eta^{\delta) (\alpha}
\eta^{\beta) [\lambda} \eta^{\rho ] \mu} + \eta^{\kappa [\lambda}
\eta^{\rho ] (\alpha} \eta^{\beta) (\gamma} \eta^{\delta) \mu}
+ \eta^{\kappa [\lambda} \eta^{\rho ] (\gamma} \eta^{\delta) (\alpha}
\eta^{\beta) \mu} \; . \qquad \label{Uvert}
\end{eqnarray}
Note that parenthesized indices are symmetrized, whereas indices enclosed
in square brackets are antisymmetrized.

\subsection{Dimensionally Regulated Result}

\begin{figure}
\includegraphics[width=4.0cm,height=3.0cm]{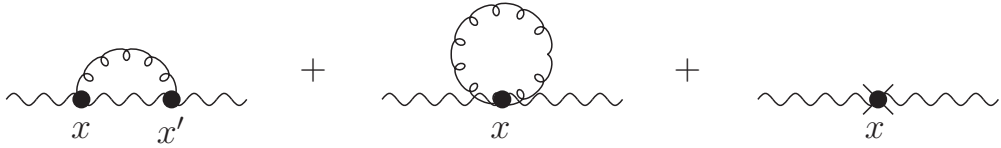}
\caption{Graviton contributions to the one loop vacuum polarization.}
\label{photon}
\end{figure}

The three one loop diagrams which contribute to the vacuum polarization
are depicted in Fig.~\ref{photon}. They can each be expressed using the
notation of the previous section. The left hand diagram is,
\begin{equation}
i \Bigl[\mbox{}^{\mu} \Pi^{\nu}_{\rm 3pt}\Bigr](x;x') = (i\kappa)^2
\partial_{\kappa} \partial_{\theta}' \Biggl\{ \!
V^{\mu\rho\kappa\lambda\alpha\beta} i\Bigl[\mbox{}_{\alpha\beta}
\Delta_{\gamma\delta}\Bigr](x;x') V^{\nu\sigma\phi\theta\gamma\delta}
\partial_{\lambda} \partial_{\phi}' i\Bigl[ \mbox{}_{\rho} \Delta_{\sigma}
\Bigr](x;x') \! \Biggr\} . \label{3pt}
\end{equation}
Substituting expressions (\ref{Gravprop}), (\ref{Photprop}) and
(\ref{Vvert}), acting the inner derivatives and performing the inner
contractions gives,
\begin{eqnarray}
\lefteqn{i \Bigl[\mbox{}^{\mu} \Pi^{\nu}_{\rm 3pt}\Bigr](x;x') = (i\kappa)^2
\frac{\Gamma^2(\frac{D}2 \!-\! 1)}{16 \pi^D} \times -(D \!-\! 3) D
\partial_{\rho} \partial_{\sigma} \Biggl\{ \frac{2 [\eta^{\mu\nu}
\eta^{\rho\sigma} \!-\! \eta^{\mu\rho} \eta^{\nu\sigma} ]}{ \Delta x^{2D-2}} }
\nonumber \\
& & \hspace{-.2cm} + \frac{D [\Delta x^{\mu} \Delta x^{\rho} \eta^{\nu\sigma}
\!-\! \Delta x^{\rho} \Delta x^{\sigma} \eta^{\mu\nu}]}{\Delta x^{2D}} -
\frac{D [\Delta x^{\mu} \Delta x^{\nu} \eta^{\rho\sigma} \!-\! \Delta x^{\rho}
\Delta x^{\nu} \eta^{\mu\sigma}]}{ \Delta x^{2D}} \Biggr\} . \qquad
\end{eqnarray}
The next step is to act the outer derivatives, at which point we
can extract a manifestly transverse form,
\begin{eqnarray}
\lefteqn{i \Bigl[\mbox{}^{\mu} \Pi^{\nu}_{\rm 3pt}\Bigr](x;x') } \nonumber \\
& & \hspace{.5cm} =  - \frac{\kappa^2 \Gamma^2(\frac{D}2 \!-\! 1)}{16 \pi^D}
(D \!-\! 3)(D \!-\! 2)^2 D \Biggl\{ \frac{(D \!+\! 1) \eta^{\mu\nu}}{\Delta
x^{2D}} \!-\! \frac{2 D \Delta x^{\mu} \Delta x^{\nu}}{\Delta x^{2D+2}}
\Biggr\} , \qquad \\
& & \hspace{.5cm} = - \frac{\kappa^2 \Gamma^2(\frac{D}2 \!-\! 1)}{16 \pi^D}
\frac{(D \!-\! 3) (D\!-\! 2)^2 D}{2 (D \!-\! 1)} \Bigl[ \eta^{\mu\nu}
\partial^2 \!-\! \partial^{\mu} \partial^{\nu} \Bigr] \frac1{\Delta x^{2D-2}}
\; . \qquad \label{transverse}
\end{eqnarray}

The middle diagram of Fig.~1 is,
\begin{equation}
i \Bigl[\mbox{}^{\mu} \Pi^{\nu}_{\rm 4pt}\Bigr](x;x') = i \kappa^2
\partial_{\kappa} \Biggl\{ U^{\mu\nu\kappa\lambda\alpha\beta\gamma\delta}
i\Bigl[\mbox{}_{\alpha\beta} \Delta_{\gamma\delta}\Bigr](x;x)
\partial_{\lambda} \delta^D(x \!-\! x') \Biggr\} . \label{4pt}
\end{equation}
This diagram vanishes because the coincidence limit of the massless
scalar propagator in flat space is zero in dimensional regularization,
$i\Delta(x;x) = 0$. The diagram on the right of Fig.~1 is,
\begin{equation}
i \Bigl[\mbox{}^{\mu} \Pi^{\nu}_{\rm ctm}\Bigr](x;x') = i 4 C_4
\Bigl( \eta^{\mu\nu} \partial^2 \!-\! \partial^{\mu} \partial^{\nu} \Bigr)
\partial^2 \delta^D(x \!-\! x') \; . \label{ctm}
\end{equation}

\subsection{Renormalization}

To renormalize (\ref{transverse}) we must first localize the
ultraviolet divergence so that it can be subtracted by the
counterterm (\ref{ctm}). This process of localization is
accomplished by first partially integrating the factor of $1/\Delta
x^{2D-2}$ in (\ref{transverse}) until the remainder is integrable
\cite{TW1}. In dimensional regularization the steps are \cite{OW},
\begin{equation}
\frac1{\Delta x^{2D-2}} = \frac{\partial^2}{2 (D \!-\! 2)^2}
\frac1{\Delta x^{2D-4}} = \frac{\partial^4}{4 (D \!-\! 2)^2
(D\!-\!3) (D\!-\!4)} \frac1{\Delta x^{2D-6}} \; . \label{IBP}
\end{equation}
Next we add zero in the form \cite{OW},
\begin{equation}
\partial^2 \frac1{\Delta x^{D-2}} = \frac{i 4 \pi^{\frac{D}2} }{
\Gamma(\frac{D}2 \!-\! 1)} \, \delta^D(x \!-\! x') \; . \label{zero}
\end{equation}
Adding (\ref{zero}) to the key part of (\ref{IBP}) in a
dimensionally consistent way gives,
\begin{eqnarray}
\lefteqn{\frac{\partial^2}{D \!-\! 4} \Biggl\{ \frac1{\Delta
x^{2D-6}} \Biggr\} } \nonumber \\
& & \hspace{1cm} = \frac{i 4 \pi^{\frac{D}2} }{\Gamma(\frac{D}2
\!-\! 1)} \frac{\mu^{D-4} \delta^D(x \!-\! x')}{D \!-\! 4} +
\frac{\partial^2}{D \!- \! 4} \Biggl\{ \frac1{\Delta x^{2D-6}} \!-\!
\frac{\mu^{D-4}}{\Delta x^{D-2}} \Biggr\} \; , \qquad \\
& & \hspace{1cm} = \frac{i 4 \pi^{\frac{D}2} }{\Gamma(\frac{D}2
\!-\! 1)} \frac{\mu^{D-4} \delta^D(x \!-\! x')}{D \!-\! 4} -
\frac{\partial^2}{2} \Biggl\{ \frac{\ln(\mu^2 \Delta x^2)}{\Delta
x^2} \Biggr\} + O( D\!-\! 4) \; . \qquad \label{localdiv}
\end{eqnarray}
Substituting (\ref{IBP}) and (\ref{localdiv}) into
(\ref{transverse}) results in the desired localized divergence,
\begin{eqnarray}
\lefteqn{ i\Bigl[\mbox{}^{\mu} \Pi^{\nu}_{\rm 3pt}\Bigr](x;x') =
-\frac{i\kappa^2 \Gamma(\frac{D}2 \!-\! 1)}{4 \pi^{\frac{D}2} }
\frac{D}{8 (D \!-\! 1) (D\!-\! 4)} \Bigl[ \eta^{\mu\nu} \partial^2
\!-\! \partial^{\mu} \partial^{\nu}\Bigr] \partial^2 \delta^D(x
\!-\! x') } \nonumber \\
& & \hspace{2.8cm} + \frac{\kappa^2}{192 \pi^4} \Bigl[ \eta^{\mu\nu}
\partial^2 \!-\! \partial^{\mu} \partial^{\nu}\Bigr] \partial^4
\Biggl\{ \frac{\ln(\mu^2 \Delta x^2)}{\Delta x^2} \Biggr\} + O(D
\!-\! 4) \; . \qquad \label{3ptexp}
\end{eqnarray}

The local divergence of expression (\ref{3ptexp}) will be completely
canceled by the counterterm (\ref{ctm}) if we make the choice,
\begin{equation}
C_4 = \frac{ \kappa^2 \Gamma(\frac{D}2 \!-\! 1)}{16 \pi^{\frac{D}2}}
\frac{D}{8 (D \!-\! 1) (D \!-\! 4)} \; . \label{C4}
\end{equation}
We can then take the unregulated limit ($D \rightarrow 4$) to obtain
the fully renormalized graviton contribution to the one loop vacuum
polarization,
\begin{equation}
\Bigl[ \mbox{}^{\mu} \Pi^{\nu}_{\rm ren}\Bigr](x;x') = -\frac{i
\kappa^2}{192 \pi^4} \Bigl[ \eta^{\mu\nu} \partial^2 \!-\!
\partial^{\mu} \partial^{\nu}\Bigr] \partial^4 \Biggl\{
\frac{\ln(\mu^2 \Delta x^2)}{\Delta x^2} \Biggr\} \; .
\label{renorm}
\end{equation}
Note that the ambiguity regarding the finite part of the counterterm
is reflected in the dimensional regularization scale $\mu$.

\section{Quantum Corrected Maxwell Equations}

The purpose of this section is to use our one result (\ref{renorm})
for the one loop vacuum polarization to quantum correct Maxwell's
equations, and then infer quantum gravitational corrections to
electrodynamics by solving these equations. We begin by deriving the
causal effective field equations of the Schwinger-Keldysh formalism.
Subsequent subsections solve these equations perturbatively for the
special cases of free photons, a point dipole pulse, an alternating
point dipole, and a static point charge.

\subsection{Schwinger-Keldysh Formalism}

We come now to the question of what to use for the vacuum polarization
$[\mbox{}^{\mu}\Pi^{\nu}](x;x')$ in the quantum corrected Maxwell
equation (\ref{qmax1}). It might seem obvious that the in-out result
(\ref{renorm}) we have just derived should be used, but that would
lead to two problems:
\begin{itemize}
\item{{\it Causality} --- The in-out vacuum polarization (\ref{renorm})
is nonzero for points $x^{\prime \mu}$ which lie in the future of
$x^{\mu}$, or at spacelike separation from it; and}
\item{{\it Reality} --- The in-out vacuum polarization (\ref{renorm})
is not real.}
\end{itemize}
One can get the right result for a static potential by simply
ignoring the imaginary part \cite{AFR,Coulomb}, but circumventing
the limitations of the in-out formalism becomes more and more
difficult as time dependent sources and higher order corrections are
included, and these techniques break down entirely for the case of
cosmology in which there may not even be asymptotic vacua. Note that
there is nothing wrong with the in-out vacuum polarization
(\ref{renorm}); it is exactly the right thing to correct the photon
propagator for asymptotic scattering computations in flat space. The
point is rather that employing (\ref{renorm}) in equation
(\ref{qmax1}) fails to provide a set of field equations with the
same scope and power as the classical Maxwell's equations.

The more appropriate field equations are those of the
Schwinger-Keldysh formalism. This technique provides a way of
computing true expectation values that is almost as simple as the
Feynman diagrams which produce in-out matrix elements
\cite{earlySK}. We shall develop the Schwinger-Keldysh rules in the
context of a scalar field $\varphi(x)$ whose Lagrangian (the space
integral of its Lagrangian density) at time $t$ is $L[\varphi(t)]$.
Suppose we are given a Heisenberg state $\vert \Psi\rangle$ whose
wave functional in terms of the operator eigenkets at time $t_0$ is
$\Psi[\varphi(t_0)]$, and we wish to take the expectation value, in
the presence of this state, of a product of two functionals of the
field operator: $A[\varphi]$, which is anti-time-ordered, and
$B[\varphi]$, which is time-ordered. The Schwinger-Keldysh
functional integral for this is \cite{FW},
\begin{eqnarray}
\lefteqn{\Bigl\langle \Psi \Bigl\vert A[\varphi] B[\varphi] \Bigr\vert \Psi
\Bigr\rangle = \Fint [d\varphi_{\scriptscriptstyle +}]
[d\varphi_{\scriptscriptstyle -}] \,
\delta\Bigl[\varphi_{\scriptscriptstyle -}(t_1) \!-\!
\varphi_{\scriptscriptstyle +}(t_1)\Bigr] } \nonumber \\
& & \hspace{1.9cm} \times A[\varphi_{\scriptscriptstyle -}]
B[\varphi_{\scriptscriptstyle +}]
\Psi^*[\varphi_{\scriptscriptstyle -}(t_0)] e^{i \int_{t_0}^{t_1} dt
\Bigl\{L[\varphi_{\scriptscriptstyle +}(t)] -
L[\varphi_{\scriptscriptstyle -}(t)]\Bigr\}}
\Psi[\varphi_{\scriptscriptstyle +}(t_0)] \; . \qquad \label{fund}
\end{eqnarray}
The time $t_1 > t_0$ is arbitrary as long as no operator in either
$A[\varphi]$ or $B[\varphi]$ is evaluated at a later time.

The Schwinger-Keldysh rules can be read off from its functional
representation (\ref{fund}). Because the same field operator is
represented by two different dummy functional variables,
$\varphi_{\scriptscriptstyle \pm}(x)$, the endpoints of lines carry
a $\pm$ polarity. External lines associated with the
anti-time-ordered operator $A[\varphi]$ have the $-$ polarity
whereas those associated with the time-ordered operator $B[\varphi]$
have the $+$ polarity. Interaction vertices are either all $+$ or
all $-$. Vertices with $+$ polarity are the same as in the usual
Feynman rules whereas vertices with the $-$ polarity have an
additional minus sign. If the state $\vert \Psi\rangle$ is something
other than free vacuum then it contributes additional interaction
vertices on the initial value surface \cite{KOW}.

Propagators can be $++$, $+-$, $-+$, or $--$. All four polarity
variations can be read off from the fundamental relation
(\ref{fund}) when the free Lagrangian is substituted for the full
one. It is useful to denote canonical expectation values in the free
theory with a subscript $0$. With this convention we see that the
$++$ propagator is just the ordinary Feynman propagator,
\begin{equation}
i\Delta_{\scriptscriptstyle ++}(x;x') = \Bigl\langle \Omega \Bigl\vert
T\Bigl(\varphi(x) \varphi(x') \Bigr) \Bigr\vert \Omega \Bigr\rangle_0 =
i\Delta(x;x') \; , \label{++}
\end{equation}
where $T$ stands for time-ordering and $\overline{T}$ denotes
anti-time-ordering. The other polarity variations are simple to read off and
to relate to the Feynman propagator,
\begin{eqnarray}
i\Delta_{\scriptscriptstyle -+}(x;x') \!\!\! & = & \!\!\! \Bigl\langle \Omega
\Bigl\vert \varphi(x) \varphi(x') \Bigr\vert \Omega \Bigr\rangle_0 \!\!=
\theta(t\!-\!t') i\Delta(x;x') \!+\! \theta(t'\!-\!t) \Bigl[i\Delta(x;x')
\Bigr]^* \! , \qquad \label{-+} \\
i\Delta_{\scriptscriptstyle +-}(x;x') \!\!\! & = & \!\!\! \Bigl\langle \Omega
\Bigl\vert \varphi(x') \varphi(x) \Bigr\vert \Omega \Bigr\rangle_0 \!\!=
\theta(t\!-\!t') \Bigl[i\Delta(x;x')\Bigr]^* \!\!+\! \theta(t'\!-\!t)
i\Delta(x;x') , \qquad \label{+-} \\
i\Delta_{\scriptscriptstyle --}(x;x') \!\!\!\!\! & = & \!\!\! \Bigl\langle
\Omega \Bigl\vert \overline{T}\Bigl(\varphi(x) \varphi(x') \Bigr) \Bigr\vert
\Omega \Bigr\rangle_0 \!\!= \Bigl[i\Delta(x;x')\Bigr]^* . \label{--}
\end{eqnarray}
In our case, both the photon and the graviton propagators depend
upon the massless scalar propagator (\ref{scalprop}), which is a
function of the Lorentz interval $\Delta x^2(x;x')$. It follows from
relations (\ref{-+}-\ref{--}) that the various Schwinger-Keldysh
propagators can be obtained by making simple replacements for the
Lorentz interval,
\begin{eqnarray}
\Delta x^2_{\scriptscriptstyle ++}(x;x') & \equiv & \Bigl\Vert
\vec{x} \!-\! \vec{x}' \Bigr\Vert^2 - c^2 \Bigl( \vert t \!-\!
t'\vert \!-\! i \epsilon
\Bigr)^2 \; , \label{Dx++} \\
\Delta x^2_{\scriptscriptstyle +-}(x;x') & \equiv & \Bigl\Vert
\vec{x} \!-\! \vec{x}' \Bigr\Vert^2 - c^2 \Bigl( t \!-\! t' \!+\!
i \epsilon\Bigr)^2 \; , \label{Dx+-} \\
\Delta x^2_{\scriptscriptstyle -+}(x;x') & \equiv & \Bigl\Vert
\vec{x} \!-\! \vec{x}' \Bigr\Vert^2 - c^2 \Bigl( t \!-\! t' \!-\!
i \epsilon\Bigr)^2 \; , \label{Dx-+} \\
\Delta x^2_{\scriptscriptstyle --}(x;x') & \equiv & \Bigl\Vert
\vec{x} \!-\! \vec{x}' \Bigr\Vert^2 - c^2 \Bigl( \vert t \!-\!
t'\vert \!+\! i \epsilon \Bigr)^2 \; . \label{Dx--}
\end{eqnarray}

Because each external line can be either $+$ or $-$ in the
Schwinger-Keldysh formalism, every 1PI N-point function of the
in-out formalism gives rise to $2^N$ 1PI N-point functions in the
Schwinger-Keldysh formalism. For every classical field $\phi(x)$ of
an in-out effective action, the corresponding Schwinger-Keldysh
effective action must depend upon two fields
--- call them $\phi_{\scriptscriptstyle +}(x)$ and
$\phi_{\scriptscriptstyle -}(x)$ --- in order to access the
appropriate 1PI function \cite{lateSK}. For the scalar paradigm we
have been considering the 1PI 2-point function as the scalar
self-mass-squared, $M^2_{\scriptscriptstyle \pm\pm}(x;x')$, and the
effective action takes the form,
\begin{eqnarray}
\lefteqn{\Gamma[\phi_{\scriptscriptstyle +},\phi_{\scriptscriptstyle -}] =
S[\phi_{\scriptscriptstyle +}] - S[\phi_{\scriptscriptstyle -}]
-\frac12 \int \!\! d^4x \! \int \!\! d^4x' } \nonumber \\
& & \times \left\{\matrix{\!
\phi_{\scriptscriptstyle +}(x) M^2_{\scriptscriptstyle ++}\!(x;x')
\phi_{\scriptscriptstyle +}(x') + \phi_{\scriptscriptstyle +}(x)
M^2_{\scriptscriptstyle +-}\!(x;x') \phi_{\scriptscriptstyle -}(x') \! \cr
\!+ \phi_{\scriptscriptstyle -}(x) M^2_{\scriptscriptstyle -+}\!(x;x')
\phi_{\scriptscriptstyle +}(x') + \phi_{\scriptscriptstyle -}(x)
M^2_{\scriptscriptstyle --}\!(x;x') \phi_{\scriptscriptstyle -}(x') \!}
\right\} + O(\phi^3_{\pm}) , \qquad
\end{eqnarray}
where $S$ is the classical action. The effective field equations are
obtained by varying with respect to $\phi_{\scriptscriptstyle +}$ and then
setting both fields equal \cite{lateSK},
\begin{equation}
\frac{\delta \Gamma[\phi_{\scriptscriptstyle +},\phi_{\scriptscriptstyle -}]
}{\delta \phi_{\scriptscriptstyle +}(x)} \Biggl\vert_{\phi_{\scriptscriptstyle
\pm} = \phi} \!\!\! = \Bigl[\partial^2 - m^2\Bigr] \phi(x)
- \! \int \! d^4x' \Bigl[M^2_{\scriptscriptstyle ++}\!(x;x') +
M^2_{\scriptscriptstyle +-}\!(x;x')\Bigr] \phi(x') + O(\phi^2) . \label{efe}
\end{equation}
The two 1PI 2-point functions we would need to quantum correct the
linearized scalar field equation are $M^2_{\scriptscriptstyle ++}\!(x;x')$
and $M^2_{\scriptscriptstyle +-}\!(x;x')$. Their sum in (\ref{efe}) gives
effective field equations which are causal in the sense that the two 1PI
functions cancel unless $x^{\prime \mu}$ lies on or within the past
light-cone of $x^{\mu}$. Their sum is also real, which neither 1PI function
is separately.

From the preceding discussion it is apparent that we wish to make
the following substitution in equation (\ref{qmax1}),
\begin{equation}
\Bigl[\mbox{}^{\mu} \Pi^{\nu}\Bigr](x;x') \longrightarrow
\Bigl[\mbox{}^{\mu}_{\scriptscriptstyle +}
\Pi^{\nu}_{\scriptscriptstyle +}\Bigr](x;x') +
\Bigl[\mbox{}^{\mu}_{\scriptscriptstyle +}
\Pi^{\nu}_{\scriptscriptstyle -}\Bigr](x;x') \; ,
\end{equation}
where we can read off the appropriate Schwinger-Keldysh vacuum
polarization from expression (\ref{renorm}),
\begin{eqnarray}
\lefteqn{\Bigl[\mbox{}^{\mu}_{\scriptscriptstyle \pm}
\Pi^{\nu}_{\scriptscriptstyle \pm}\Bigr](x;x') = - \frac{(\pm)(\pm)
i \kappa^2}{192 \pi^4} \Bigl[ \eta^{\mu\nu}
\partial^2 \!-\! \partial^{\mu}
\partial^{\nu}\Bigr] \partial^4 \Biggl\{ \frac{\ln(\mu^2 \Delta
x^2_{\scriptscriptstyle \pm\pm}) }{\Delta x^2_{\scriptscriptstyle
\pm\pm} } \Biggr\} \; ,} \\
& & \hspace{.7cm}= -\frac{(\pm)(\pm) i \kappa^2}{1536 \pi^4} \Bigl[
\eta^{\mu\nu} \partial^2 \!-\! \partial^{\mu}
\partial^{\nu}\Bigr] \partial^6 \Bigl\{\ln^2(\mu^2 \Delta
x^2_{\scriptscriptstyle \pm\pm}) \!-\! 2 \ln(\mu^2 \Delta
x^2_{\scriptscriptstyle \pm\pm}) \Bigr\} \; . \qquad
\end{eqnarray}
Now define the temporal and spatial intervals as,
\begin{equation}
\Delta t \equiv t \!-\! t' \qquad , \qquad \Delta r \equiv \Vert
\vec{x} \!-\! \vec{x}' \Vert \; .
\end{equation}
It is apparent from expressions (\ref{Dx++}-\ref{Dx+-}) that
differences of logarithms of the the $++$ and $+-$ intervals give,
\begin{eqnarray}
\ln(\mu^2 \Delta x^2_{\scriptscriptstyle ++}) - \ln(\mu^2 \Delta
x^2_{\scriptscriptstyle +-}) & = & 2 \pi i \theta\Bigl(\Delta t \!-\!
\Delta r\Bigr) \; , \\
\ln^2(\mu^2 \Delta x^2_{\scriptscriptstyle ++}) - \ln^2(\mu^2 \Delta
x^2_{\scriptscriptstyle +-}) & = & 4\pi i \theta\Bigl(\Delta t \!-\!
\Delta r\Bigr) \ln\Bigl[\mu^2 (\Delta t^2 \!-\! \Delta r^2)\Bigr] \; . \qquad
\end{eqnarray}
Hence the vacuum polarization which belongs in equation
(\ref{qmax1}) is,
\begin{eqnarray}
\lefteqn{\Bigl[\mbox{}^{\mu} \Pi^{\nu}\Bigr](x;x') =
\frac{\kappa^2}{384 \pi^3} \Bigl[ \eta^{\mu\nu} \partial^2 \!-\!
\partial^{\mu} \partial^{\nu}\Bigr] } \nonumber \\
& & \hspace{3cm} \times \partial^6 \Biggl\{ \theta\Bigl(\Delta t \!-\!
\Delta r\Bigr) \Biggl[ \ln\Bigl[\mu^2 \Delta t^2 \!-\! \Delta r^2)\Bigr]
\!-\! 1\Biggr] \Biggr\} + \mathcal{O}(\kappa^4) \; . \qquad \label{finalform}
\end{eqnarray}

\subsection{Photons}

Expression (\ref{finalform}) gives all the one loop contributions which
derive exclusively from interaction vertices, but there are also
contributions from perturbative corrections to the initial state wave
functionals. In the scalar functional integral (\ref{fund}) these
wave functionals are $\Psi[\varphi_{\scriptscriptstyle +}(t_0)]$ and
$\Psi^*[\varphi_{\scriptscriptstyle -}(t_0)]$; for gravity plus
electromagnetism they would be functionals of $A_{\mu}$ and $h_{\mu\nu}$,
evaluated at the initial time.

Each state wave functional can be expressed as the wave functional of
free vacuum times a series of perturtbative corrections,
\begin{equation}
\Psi[A,h] = \Psi_0[A,h] \times \Biggl\{1 + \mathcal{O}\Bigl(\kappa h A^2
\Bigr) \Biggr\} \; .
\end{equation}
It is straightforward to show that the free vacuum contribution is
what fixes the real part of the propagator in the functional formalism
\cite{TW2}. If there were no perturbative state corrections then merely
employing the correct propagators would completely account for the
state wave functionals. However, there must be perturbative state
corrections because free vacuum cannot be the true vacuum state of an
interacting quantum field theory.

Perturbative state corrections manifest as new interactions on the
initial value surface \cite{FW}. When the initial value surface is
in the asymptotic past (or the asymptotic past and future for in-out
matrix elements) these interactions have no effect on operators at
finite times. However, they can be important when the initial value
surface is at a finite time, as it must be in cosmology. The first
correction relevant for a massless, minimally coupled $\lambda
\phi^4$ theory has recently been worked out on de Sitter background
\cite{KOW}. In this case the initial state correction is necessary
to make the linearized effective field equation well defined at the
initial time \cite{BOW}, and to eliminate an infinite series of
rapidly redshifting terms from the two loop expectation value of the
stress tensor \cite{OW}.

We shall assume that the missing state corrections exactly cancel
the surface terms which arise when (\ref{finalform}) is partially
integrated. To see what this entails, first note that all orders of
the ``pure-vertex'' part of the vacuum polarization take the
manifestly transverse form,
\begin{equation}
\Bigl[ \mbox{}^{\mu} \Pi^{\nu}\Bigr](x;x') = \Bigl( \eta^{\mu\nu}
\partial^2 \!-\! \partial^{\mu} \partial^{\nu} \Bigr) \Pi(x \!-\!
x') \; . \label{generalform}
\end{equation}
The partial integration we have in mind concerns the quantum
correction to Maxwell's equation,
\begin{eqnarray}
\lefteqn{\int \!\! d^4x' \, \Bigl(\eta^{\mu\nu} \partial^2 \!-\!
\partial^{\mu} \partial^{\nu} \Bigr) \Pi(x \!-\! x') A_{\nu}(x') }
\nonumber \\
& & \hspace{4cm} = \int \!\! d^4x' \, \Pi(x \!-\! x')
\partial_{\nu}' F^{\nu\mu}(x') + {\rm Surface\ Terms} \; . \qquad
\end{eqnarray}
In the Schwinger-Keldysh formalism the $++$ and $+-$ contributions
exactly cancel on the future temporal surface, as well as on the
surface at spatial infinity. Hence the only surface terms come from
the initial time. Of course this is also true of perturbative state
corrections. We assume that the two contributions exactly cancel, so
that the full, quantum-corrected Maxwell equation is,
\begin{equation}
\partial^{\nu} F_{\nu\mu}(x) + \int \!\! d^4x' \, \Pi(x \!-\! x') \,
\partial^{\prime \nu} \! F_{\nu\mu}(x') = J_{\mu}(x) \; . \label{qmax2}
\end{equation}

We are finally ready to consider the case of free photons, which
corresponds to $J_{\mu}(x) = 0$. Note from equation (\ref{qmax2})
that these obey $\partial^{\nu} F_{\nu\mu}(x) = 0$, the same as in
the classical theory. One might worry about the potential for
solutions of the form $\partial^{\nu} F_{\nu\mu}(x) = S_{\mu}(x)$,
where $S_{\mu}(x)$ obeys the integral equation,
\begin{equation}
S_{\mu}(x) + \int \!\! d^4x' \, \Pi(x \!-\! x') S_{\mu}(x') = 0 \; .
\end{equation}
However, an effective field equation such as (\ref{qmax2}) can only
be used to perturbatively correct classical solutions \cite{Simon},
which means we must exclude any such solutions. Hence we conclude
that {\it quantum gravity on flat space background makes no
correction to free photons at any order}, except for possible field
strength renormalization.

\subsection{Instantaneously Creating A Point Dipole}

The charge density of a static point electric dipole $\vec{p}$ at
the origin is $\rho = -\vec{p} \cdot \vec{\nabla} \delta^3(\vec{x})$.
We might imagine creating such a dipole at the instant $t=0$ by
separating the charges in a very small, neutral particle such as a
neutron. The conserved 4-current associated with this event is,
\begin{equation}
J^0(t,\vec{x}) = -\theta(t) \vec{p} \!\cdot\! \vec{\nabla}
\delta^3(\vec{x}) \qquad , \qquad J^i(t,\vec{x}) = p^i \delta(t)
\delta^3(\vec{x}) \; . \label{J4Idipole}
\end{equation}
The response of the magnetic field provides a good perturbative
illustration of the smearing of the light-cone which was conjectured
so long ago \cite{smear}.

Before proceeding it is desirable to reorganize equation (\ref{qmax2})
in two ways. The first has to do with the limitation inherent in only
possessing the first order term in the loop expansion of $\Pi(x-x')$,
\begin{equation}
\Pi(x \!-\! x') = \Pi^{(1)}(x \!-\! x') + \Pi^{(2)}(x \!-\! x') +
\mathcal{O}(\kappa^6) \; . \label{Piexp}
\end{equation}
Of course this means we can only infer the one loop correction to
the field strength, so we may as well expand it,
\begin{equation}
F_{\mu\nu}(x) = F^{(0)}_{\mu\nu}(x) + F^{(1)}_{\mu\nu}(x) +
F^{(2)}_{\mu\nu}(x) + \mathcal{O}(\kappa^6) \; . \label{Fexp}
\end{equation}
Substituting (\ref{Piexp}) and (\ref{Fexp}) in the quantum-corrected
Maxwell equation (\ref{qmax2}) and segregating different orders of
$\kappa^2$ produces the hierarchy,
\begin{eqnarray}
\partial^{\nu} F^{(0)}_{\nu\mu}(x) & = & J_{\mu}(x) \; , \label{0th} \\
\partial^{\nu} F^{(1)}_{\nu\mu}(x) & = & -\int \!\! d^4x' \,
\Pi(x \!-\! x') J_{\mu}(x') \equiv J^{(1)}_{\mu}(x) \; , \qquad
\label{1st}
\end{eqnarray}
and so on. Note the classical source $J_{\mu}(x)$ is 0th order.

The second reorganization concerns deriving the field strength
directly, without constructing the vector potential. Consider
taking the curl of the classical Maxwell equation,
\begin{equation}
\epsilon^{\rho\sigma \mu\nu} \partial_{\mu} \partial^{\alpha}
F_{\alpha \nu} = \partial^2 \epsilon^{\rho\sigma\mu\nu} \partial_{\mu}
A_{\nu} = \epsilon^{\rho\sigma\mu\nu} \partial_{\mu} J_{\nu}
\quad \Longrightarrow \quad \partial^2 F_{\mu\nu} =
\partial_{\mu} J_{\nu} - \partial_{\nu} J_{\mu} \; .
\end{equation}
Combining this with (\ref{0th}-\ref{1st}) implies,
\begin{eqnarray}
\partial^2 F^{(0)}_{\mu\nu} & = & \partial_{\mu} J_{\nu} -
\partial_{\nu} J_{\mu} \; , \label{cmax} \\
\partial^2 F^{(1)}_{\mu\nu} & = & \partial_{\mu} J^{(1)}_{\nu} -
\partial_{\nu} J^{(1)}_{\mu} \; . \label{qmax3}
\end{eqnarray}
Now recall that our one loop current density can be expressed
as the d'Alem\-ber\-tian of something,
\begin{eqnarray}
J^{(1)}_{\mu}(x) & \equiv & -\int \!\! d^4x' \, \Pi^{(1)}(x \!-\! x')
J_{\mu}(x') \; , \\
& = & \frac{i \kappa^2 \partial^4}{192 \pi^4} \int \!\! d^4x' \,
\Biggl\{ \frac{ \ln(\mu^2 \Delta x^2_{\scriptscriptstyle ++})}{
\Delta x^2_{\scriptscriptstyle ++}} \!-\! \frac{\ln(\mu^2 \Delta x^2_{
\scriptscriptstyle +-})}{\Delta x^2_{\scriptscriptstyle +-}} \Biggr\}
J_{\mu}(x') \; , \qquad \label{form1} \\
& = & \frac{\kappa^2 \partial^6}{384 \pi^3} \int \!\! d^4x' \,
\theta\Bigl(\Delta t \!-\! \Delta r\Bigr) \Biggl\{ \ln\Bigl[ \mu^2
\Bigl( \Delta t^2 \!-\! \Delta r^2\Bigr) \Bigr] \!-\! 1\Biggr\}
J_{\mu}(x') \; . \qquad \label{form2}
\end{eqnarray}
Comparison of (\ref{qmax3}) with (\ref{form1}) or (\ref{form2}) implies
a result for the one loop field strength, up to possible homogeneous
terms,
\begin{eqnarray}
F^{(1)}_{\mu\nu}(x) & \!\!=\!\! & \frac{i \kappa^2 \partial^2}{192 \pi^4} \,
2\partial_{[\mu} \!\! \int \!\! d^4x' \, \Biggl\{ \frac{ \ln(\mu^2 \Delta
x^2_{\scriptscriptstyle ++})}{\Delta x^2_{\scriptscriptstyle ++}} \!-\!
\frac{\ln(\mu^2 \Delta x^2_{\scriptscriptstyle +-})}{\Delta x^2_{
\scriptscriptstyle +-}} \Biggr\} J_{\nu]}(x') \; , \qquad \label{qmax4a} \\
&\!\! = \!\!& \frac{\kappa^2 \partial^4}{384 \pi^3} \, 2 \partial_{[\mu} \!\!
\int \!\! d^4x' \, \theta\Bigl(\Delta t \!-\! \Delta r\Bigr) \Biggl\{ \ln\Bigl[
\mu^2 \Bigl( \Delta t^2 \!-\! \Delta r^2\Bigr) \Bigr] \!-\! 1\Biggr\}
J_{\nu]}(x') \; . \qquad \label{qmax4b}
\end{eqnarray}

We are now ready to specialize to the current density (\ref{J4Idipole})
of an instantaneously created dipole. Substituting in (\ref{0th}) and
specializing to purely spatial indices gives,
\begin{equation}
\partial^2 F^{(0)}_{ij}(x) = \Bigl(\partial_i p_j \!-\! \partial_j p_i\Bigr)
\delta^4(x) \; .
\end{equation}
The solution can be expressed in a convenient form by noting the
$D=4$ dimensional version of relation (\ref{zero}),
\begin{equation}
\partial^2 \Biggl\{\frac1{\Delta x^2_{\scriptscriptstyle ++}} \Biggr\}
= 4 \pi^2 i \delta^4(x \!-\! x') \qquad , \qquad
\partial^2 \Biggl\{ \frac1{\Delta x^2_{\scriptscriptstyle +-}} \Biggr\}
= 0 \; .
\end{equation}
Hence we have,
\begin{equation}
F^{(0)}_{ij}(x) = -\frac{i}{4 \pi^2} \Bigl( \partial_i p_j \!-\! \partial_j
p_i\Bigr) \Biggl\{ \frac1{\Delta x^2_{\scriptscriptstyle ++}} \!-\!
\frac1{\Delta x^2_{\scriptscriptstyle +-}} \Biggr\} \; ,
\end{equation}
where ${x'}^{\mu} = 0$ is understood. Now write out the two intervals,
\begin{eqnarray}
\Delta x^2_{\scriptscriptstyle ++} & = & r^2 - t^2 + \epsilon^2 + 2
\epsilon \vert t\vert i \; , \\
\Delta x^2_{\scriptscriptstyle +-} & = & r^2 - t^2 + \epsilon^2 - 2
\epsilon t i \; .
\end{eqnarray}
Combining these relations with the Dirac identity results in the familiar
form for the Li\'enard-Wiechert potential,
\begin{equation}
F^{(0)}_{ij}(x) = -\frac1{2 \pi} \Bigl( \partial_i p_j \!-\! \partial_j
p_i\Bigr) \theta(t) \delta(r^2 \!-\! t^2) \; . \label{B0}
\end{equation}

The most convenient form for the one loop correction is (\ref{qmax4a}),
\begin{eqnarray}
F^{(1)}_{ij}(x) & = & \Bigl( \partial_i p_j \!-\! \partial_j p_i\Bigr)
\frac{i \kappa^2 \partial^2}{192 \pi^4} \Biggl\{ \frac{ \ln(\mu^2 \Delta
x^2_{\scriptscriptstyle ++})}{\Delta x^2_{\scriptscriptstyle ++}} \!-\!
\frac{\ln(\mu^2 \Delta x^2_{\scriptscriptstyle +-})}{\Delta x^2_{
\scriptscriptstyle +-}} \Biggr\} \; , \qquad \\
& = & \Bigl( \partial_i p_j \!-\! \partial_j p_i\Bigr)
\frac{i \kappa^2 \partial^2}{48 \pi^4} \Biggl\{
\frac1{\Delta x^4_{\scriptscriptstyle ++}} \!-\!
\frac1{\Delta x^4_{\scriptscriptstyle +-}} \Biggr\} \; , \qquad \\
 & = & \Bigl( \partial_i p_j \!-\! \partial_j p_i\Bigr) \Bigl(
\frac{\kappa^2}{12 \pi^2} \frac{\partial}{\partial r^2} \Bigr)
\Biggl\{ \frac{\theta(t) \delta(r^2 \!-\! t^2)}{2\pi} \Biggr\} \; .
\label{B1}
\end{eqnarray}
Adding the one loop magnetic field (\ref{B1}) to the tree one (\ref{B0})
leads to an interesting form,
\begin{eqnarray}
F_{ij}(x) & = & -\frac1{2 \pi} \Bigl( \partial_i p_j \!-\! \partial_j p_i\Bigr)
\theta(t) \Biggl\{ 1 \!-\! \frac{\kappa^2}{12 \pi^2} \frac{\partial}{\partial
r^2} \Biggr\} \delta(r^2 \!-\! t^2) + \mathcal{O}(\kappa^4) \; , \\
& = & -\frac1{2 \pi} \Bigl( \partial_i p_j \!-\! \partial_j
p_i\Bigr) \theta(t) \delta\Bigl(r^2 \!-\! t^2 \!-\!
\frac{\kappa^2}{12 \pi^2}\Bigr) + \mathcal{O}(\kappa^4) \; . \qquad
\label{smeared}
\end{eqnarray}
It would therefore be fair to say that, by time $t$ the signal has
reached a distance $r$ slightly outside the classical light-cone,
\begin{equation}
r^2 = t^2 + \frac{\kappa^2}{12 \pi^2} + \mathcal{O}(\kappa^4) \; .
\label{newcone}
\end{equation}

Although intriguing, the super-luminality we have just found is
unobservably small. In particular, it cannot serve as any sort
of explanation for the OPERA result \cite{OPERA}. It also isn't
cumulative, so looking at cosmological sources makes the effect 
no larger. Another thing our effect fails to do is break Lorentz
invariance, as could have been predicted from the fact that 
perturbative quantum gravity provides no mechanism for spontaneously 
breaking this symmetry. Instead of signals propagating along the
classical light-cone $\eta_{\mu\nu} x^{\mu} x^{\nu} = 0$, they 
now propagate along $\eta_{\mu\nu} x^{\mu} x^{\nu} = \frac{4 G}{3\pi}$. 
So it is not that the speed of light or the dispersion relation
has been changed. 

Of course theories with a nonlinear kinetic operator can show
super-luminal propagation even classically \cite{BMV}. Our effect is
different in that it arises from quantum fluctuations of the metric
operator which sets the light-cone. One interpretation for the net
super-luminal propagation is that there is more volume outside the
classical light-cone than inside. This might be checked by extending 
the graviton expansion of the volume of the past light-cone one order 
higher than in \cite{PW4} and then computing its expectation value. 
If the one loop correction is positive then our conjecture is verified. 
Note also that this check would be independent of the choice of gauge 
because the volume of the past light-cone is a gauge invariant operator.

There have been many claims of super-luminal propagation from quantum
electrodynamics in nontrivial geometries \cite{DH,DS}. Our result is
different in that it occurs in flat space, and is due to fluctuations
of the metric operator, rather than of some matter field. We also
doubt that the earlier claims result from true super-luminal propagation.
One cannot compute the vacuum polarization produced by fermions in an 
arbitrary geometry because the fermion propagator is not known for general
metric. What was done instead is a derivative expansion. This should
be valid for low energy effective field theory; for example, it should
give correct results for the phase velocity of some continuous, low
frequency signal. However, demonstrating true super-luminality requires 
following the propagation of a pulse, and the high frequency modes which
are essential for this are not correctly treated by derivative expansions. 
In fact the Schwinger-Keldysh formalism \cite{earlySK} implies there 
cannot be super-luminal propagation from the fermionic contribution to 
vacuum polarization.

\subsection{An Alternating Point Dipole}

The 4-current associated with an alternating point dipole is,
\begin{equation}
\label{eq:current}
J^0(t,\vec{x}) = - \vec{p} \!\cdot\! \vec{\nabla}
\delta^3(\vec{x})e^{-i\omega t} \qquad , \qquad 
J^i(t,\vec{x}) = -i \omega p^i \delta^3(\vec{x}) e^{-i\omega t} \; .
\end{equation}
To find the quantum correction to the current we employ the same 
expansion technique used in the previous section where the first 
order correction is defined as,
\begin{equation}
\label{eq:int}
J^\mu_{(1)}(x)= - \frac{G\partial^6}{24\pi^2} \!\! \int \!\! d^4x' \, 
\theta(\Delta t-\Delta x) \Bigl\{ \ln\Bigl[\mu^2 (\Delta t^2 \!-\!
\Delta x^2)\Bigr] \!-\! 1 \Bigr\} J^\mu(x') \; .
\end{equation}
We can evaluate this integral by rewriting $x$ and the differential 
operators as $x = \frac{1}{2} (x+t) + \frac{1}{2} (x-t)$ and 
$\partial^2 = \frac{1}{x} (\partial_x-\partial_t) (\partial_x+\partial_t) x$. 
Thus we come to the convienient form,
\begin{equation}
\label{eq:diff}
\partial^4 = \frac{1}{2x} \Bigl(\partial_x \!-\! \partial_t\Bigr)^2 
\Bigl(\partial_x \!+\!\partial_t\Bigr)^2 (x \!+\! t) \!+\!
\frac{1}{2x} \Bigl(\partial_x \!-\! \partial_t\Bigr)^2 
\Bigl(\partial_x \!+\! \partial_t\Bigr)^2 (x\!-\!t) \; .
\end{equation}
By subsituting (\ref{eq:diff}) for $\partial^4$ in (\ref{eq:int}) and 
applying the zeroth order currents (\ref{eq:current}) we find the one loop 
currents to be,
\begin{eqnarray}
J^0_{(1)}(t,\vec{x}) & = & \partial^2 \Biggl\{ \frac{G \vec{p} \!\cdot\!
\vec{\nabla}}{6\pi^2} \Bigl[ -\frac{i \omega}{x^2} \!+\! \frac1{x^3}
\Bigr] e^{-i\omega(t-x)} \Biggr\} \; , \label{rho1dip}  \\
J^i_{(1)}(t,\vec{x}) & = & \partial^2 \Biggl\{ \frac{i \omega p^i G}{6\pi^2} 
\Bigl[ -\frac{i \omega}{x^2} \!+\! \frac1{x^3} \Bigr] 
e^{-i\omega(t-x)} \Biggr\} \; . \label{J1dip}
\end{eqnarray}

From (\ref{cmax}) we see that the zeroth order field strengths for this 
source obey,
\begin{eqnarray}
\partial^2 F_{0i}^{(0)}(t,\vec{x}) & = & -\Bigl[\omega^2 p_i \!-\! 
\partial_i\vec{p} \!\cdot\! \vec{\nabla} \Bigr] \, \delta^3(\vec{x}) 
e^{-i\omega t} \; , \\
\partial^2 F_{ij}^{(0)}(t,\vec{x} & = & -i\omega \Bigl[\partial_i p_j \!-\!
\partial_j p_i \Bigr] \, \delta^3(\vec{x}) e^{-i\omega t} \; .
\end{eqnarray}
Applying the Li\'enard-Wiechert Green's function we find,
\begin{eqnarray}
F_{0i}^{(0)}(t,\vec{x}) & = & \frac{1}{4\pi} [\omega^2 p_i \!-\! 
\partial_i \vec{p} \!\cdot\! \vec{\nabla}\Bigr] \, \frac{e^{-i\omega(t-x)}}{x} 
\; , \\
F_{ij}^{(0)}(t,\vec{x}) & = & \frac{i\omega}{4\pi} \Bigl[\partial_i p_j 
\!-\! \partial_j p_i \Bigr] \, \frac{e^{-i\omega(t-x)}}{x} \; .
\end{eqnarray}
From (\ref{qmax3}) we see that one loop field strengths follow by simply
deleting the $\partial^2$ from (\ref{rho1dip}-\ref{J1dip}) and acting 
some derivatives,
\begin{eqnarray}
F_{0i}^{(1)}(t,\vec{x}) & = & -\frac{i\omega G}{6\pi^2} \Bigl[\omega^2 p_i
\!-\! \partial_i \vec{p} \!\cdot\! \vec{\nabla}\Bigr] \Bigl[1 \!+\!
\frac{i}{\omega x}\Bigr] \, \frac{e^{-i\omega(t-x)}}{x^2} \; , \\
F_{ij}^{(1)}(t,\vec{x}) & = & -\frac{i\omega G}{6\pi^2}(i\omega) 
\Bigl[\partial_i p_j \!-\! \partial_j p_i\Bigr] \Bigl[1 \!+\! 
\frac{i}{\omega x} \Bigr] \, \frac{e^{-i\omega(t-x)}}{x^2} \; .
\end{eqnarray}
Adding the loop correction to the tree results gives,
\begin{eqnarray}
F_{0i}(t,\vec{x}) & \!\!\!\!=\!\!\!\! & \Bigl[\omega^2 p_i \!-\!
\partial_i \vec{p} \!\cdot\! \vec{\nabla}\Bigr] \!\times\! 
\frac{e^{-i\omega(t-x)}}{4 \pi x} \!\times\! \Biggl\{1 \!-\! 
\frac{2i\omega G}{3\pi x} \Bigl[1 \!+\! \frac{i}{\omega x}\Bigr] \!+\! 
\mathcal{O}(G^2) \Biggr\} \; , \qquad \\
F_{ij}(t,\vec{x}) & \!\!\!\!=\!\!\!\! & i\omega \Bigl[\partial_i p_j \!-\!
\partial_j p_i\Bigr] \times \frac{e^{-i\omega(t-x)}}{4 \pi x} \times 
\Biggl\{1 \!-\! \frac{2i\omega G}{3\pi x} \Bigl[1 \!+\! \frac{i}{\omega x}
\Bigr] \!+\! \mathcal{O}(G^2) \Biggr\} \; . \qquad
\end{eqnarray}
Of course the obvious conclusion is that the one loop corrections have
no effect in the far field regime, and the near field regime is 
unobservably close to the source.

\subsection{A Static Point Charge}

The charge density of a static point charge $q$ at the origin is 
$\rho = q \delta^3(\vec{x})$. The conserved 4-current associated with 
this source is,
\begin{equation}
J^\mu(x) = q \delta^3(\vec{x}) \delta^\mu_0 \; .
\end{equation}
Because the $\mu = 0$ component differs from the alternating dipole
of the previous subsection only by setting $\omega = 0$ and replacing
$-\vec{p} \!\cdot\! \vec{\nabla}$ with $q$, we can read off the one 
loop current density from (\ref{rho1dip}),
\begin{equation}
J^0_{(1)}(t,\vec{x}) = -\frac{Gq}{\pi^2x^5} \; .
\end{equation}
Of course the vector components vanish so we find the correction to
the Coulomb potential is,
\begin{equation}
\Phi(r) = \frac{q}{4 \pi r} \Biggr\{ 1 + \frac{2 G}{3\pi r^2} 
+ \mathcal{O}(G^2) \Biggr\} \; . \label{Cpot}
\end{equation}
 
\begin{figure}

\begin{center}
\includegraphics[width=3.0cm,height=3.0cm]{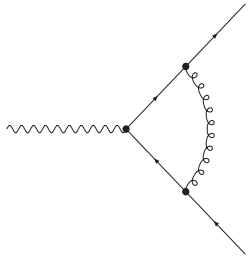}
\end{center}

\caption{Vertex correction included (along with many other diagrams)
in the Bjerrum-Bohr result \cite{Coulomb}, but not in either our 
result (\ref{Cpot}) or that of Radkowski \cite{AFR}. Charged scalar 
lines are solid with an arrow, photon lines are wavy and graviton 
lines are winding.} 

\label{vertex}
\end{figure}

Our result (\ref{Cpot}) agrees with that found in 1970 by Radkowski 
\cite{AFR}. The one loop correction that Bjerrum-Bohr inferred from 
the scattering of charged, gravitating scalars differs from what we 
got by a factor of nine \cite{Coulomb}. Part of this discrepancy may
be due to different sources; Bjerrum-Bohr considered a charged scalar 
whereas we used a point particle with worldline $\chi^{\mu}(\tau)$,
\begin{equation}
L_{\rm point} = -m \sqrt{-g_{\mu\nu}\Bigl(\chi(\tau)\Bigr)
\dot{\chi}^{\mu}(\tau) \dot{\chi}^{\nu}(\tau) } + q \dot{\chi}^{\mu}(\tau) 
A_{\mu}\Bigl(\chi(\tau)\Bigr) \; .
\end{equation}
However, we believe the largest part of the discrepancy arises from 
Bjerrum-Bohr having implicitly included corrections to the current 
density like the diagram depicted in Fig.~\ref{vertex}. We could have
and should have done this, but we will see in the next section that
it would only have altered all of our one loop field strengths by an 
overall constant.

Radkowski \cite{AFR}, Bjerrum-Bohr \cite{Coulomb} and we all agree 
that quantum gravity strengthens the electromagnetic force at one loop. 
The opposite conclusion seems to arise from computations of the quantum 
gravitational contribution to the electromagnetic beta function 
\cite{RW,Reuter1,Reuter2}. These show that quantum gravity decreases 
the electromagnetic coupling constant at high energy scales. That would 
normally be assumed to mean that quantum gravity weakens the electromagnetic 
force at short distances, but it is well to keep in mind that the beta 
function is not directly observable. The observable thing is the strength 
scattering between charged particles, and the Bjerrum-Bohr computation 
shows that one loop quantum gravity effects weaken this, rather than 
strengthening it.

\section{Gauge Dependence}

The purpose of this section is to examine how the results of the
previous section depend upon our choices of the gravitational
gauge fixing term (\ref{GRfix}) and the electromagnetic gauge fixing 
term (\ref{EMfix}). We begin with some general considerations which
reduce the issue to a single proportionality constant. The graviton
and photon propagators are then worked out for a general 3-parameter
family of covariant gauges. Although one of these parameters drops
out, the other two can change the proportionality constant all the
way from minus infinity to plus infinity. We close by exploiting 
the gauge independent result of Bjerrum-Bohr to argue that this
seeming gauge dependence may cancel out if quantum gravitational
corrections to the current density are included.

\subsection{General Considerations}

Note from expression (\ref{3pt}) that the vacuum polarization is
transverse on each of its two indices as a trivial consequence of
the antisymmetry of the vertex function on its first and third
indices,
\begin{equation}
V^{\mu\rho\kappa\lambda\alpha\beta} = -
V^{\kappa\rho\mu\lambda\alpha\beta} \; .
\end{equation}
This is completely without regard to the gauges employed to define
graviton and photon propagators. Suppose we now restrict attention
to gauges which preserve Poincar\'e invariance. Because the 
Lagrangians (\ref{LGR}-\ref{LBPHZ}) and the background are also 
Poincar\'e invariant, the vacuum polarization must inherit this
symmetry. Then dimensional analysis, transversality and the standard 
$\kappa^2$ of a one loop quantum gravity result, together imply a 
form like that of (\ref{transverse}),
\begin{equation}
i\Bigl[\mbox{}^{\mu} \Pi^{\nu}_{\rm 3pt}\Bigr](x;x') = -{\rm
Constant} \times \kappa^2 \Bigl[\eta^{\mu\nu} \partial^2 \!-\!
\partial^{\mu} \partial^{\nu}\Bigr] \frac1{\Delta x^{2D-2}} \; .
\label{genform}
\end{equation}
However, the constant prefactor might be gauge dependent, and that
same gauge dependent constant would multiply all of our one loop
corrections.

It is useful to begin at a somewhat earlier point. If different --- 
but Poincar\'e invariant --- graviton and photon propagators had been 
employed in expression (\ref{3pt}), then the combination of Poincar\'e
invariance, dimensional analysis and the algebraic symmetries of the
vertex function and the propagators imply that the result can be 
expressed in terms of two constants $A$ and $B$,
\begin{eqnarray}
\lefteqn{(i \kappa)^2 \partial_{\kappa} \partial_{\theta}' \Biggl\{ 
V^{\mu\rho\kappa\lambda\alpha\beta} i\Bigl[\mbox{}_{\alpha\beta}
\Delta^{\rm new}_{\gamma\delta}\Bigr](x;x') V^{\nu\sigma\phi\theta\gamma\delta}
\partial_{\lambda} \partial_{\phi}' i\Bigl[ \mbox{}_{\rho} 
\Delta^{\rm new}_{\sigma} \Bigr](x;x') \Biggr\} } \nonumber \\
& & \hspace{-.5cm} = (i \kappa)^2 \frac{\Gamma^2(\frac{D}2 \!-\! 1)}{16 \pi^D} 
\, (D \!-\! 2) \partial_{\kappa} \partial_{\theta}' \Biggl\{ A \times 
\frac{4 \Delta x^{[\mu} \eta^{\kappa] [\nu} \Delta x^{\theta]}}{\Delta x^{2D}} 
+ B \times \frac{\eta^{\mu [\nu} \eta^{\theta] \kappa} }{\Delta x^{2D - 2}} 
\Biggr\} \; , \qquad \label{ABdef} \\
& & \hspace{-.5cm} = - \frac{\kappa^2 \Gamma^2(\frac{D}2 \!-\! 1)}{16 \pi^D}
\times (D \!-\! 2) \Bigl[ D A \!-\! 2 (D \!-\! 1) B\Bigr] \nonumber \\
& & \hspace{4cm} \times \Biggl\{ \frac{ (D \!+\! 1) \eta^{\mu\nu} }{\Delta 
x^{2D}} - \frac{2 D \Delta x^{\mu} \Delta x^{\nu}}{\Delta x^{2D + 2}} \Biggr\} 
\; , \qquad \\
& & \hspace{-.5cm} = - \frac{\kappa^2 \Gamma^2(\frac{D}2 \!-\! 1)}{16 \pi^D}
\times \frac{(D \!-\! 2) [D A \!-\! 2 (D \!-\! 1) B]}{2 (D \!-\! 1)} \times
\Bigl[ \eta^{\mu\nu} \partial^2 \!-\! \partial^{\mu} \partial^{\nu} \Bigr]
\frac1{\Delta x^{2 D - 2}} \; . \qquad
\end{eqnarray}
We can therefore identify the proportionality constant in 
(\ref{genform}) as,
\begin{eqnarray}
{\rm Constant} & = &
\frac{\Gamma^2(\frac{D}2 \!-\! 1)}{16 \pi^D} \times \frac{(D \!-\! 2) 
[D A \!-\! 2 (D \!-\! 1) B]}{2 (D \!-\! 1)} \; , \qquad \\
& \equiv &  
\frac{\Gamma^2(\frac{D}2 \!-\! 1)}{16 \pi^D} \times \frac12 \Bigl(
\frac{D \!-\! 2}{D \!-\! 1} \Bigr) \times C \; . \label{Cdef}
\end{eqnarray}

\subsection{General Covariant Gauges}

\begin{table}

\vbox{\tabskip=0pt \offinterlineskip
\def\tablerule{\noalign{\hrule}}
\halign to390pt {\strut#& \vrule#\tabskip=1em plus2em&
\hfil#\hfil& \vrule#& \hfil#\hfil& \vrule#&
\hfil#\hfil& \vrule#\tabskip=0pt\cr
\tablerule
\omit&height4pt&\omit&&\omit&&\omit&\cr
\omit&height2pt&\omit&&\omit&&\omit&\cr
&& $\!\!\!\! I \!\!\!\!$ && $\!\!\!\! \mathcal{C}_I(D,a,b) \!\!\!\!$
&& $\!\!\!\! [\mbox{}_{\mu\nu} \mathcal{T}^I_{\rho\sigma}] \!\!\!\!$ & \cr
\omit&height4pt&\omit&&\omit&&\omit&\cr
\tablerule
\omit&height2pt&\omit&&\omit&&\omit&\cr
&& 1 && 1 && $2 \eta_{\mu (\rho} \eta_{\sigma) \nu}$ & \cr
\omit&height2pt&\omit&&\omit&&\omit&\cr
\tablerule
\omit&height2pt&\omit&&\omit&&\omit&\cr
&& 2 && $-\frac2{D - 2}$ && $\eta_{\mu \nu} \eta_{\rho \sigma}$ & \cr
\omit&height2pt&\omit&&\omit&&\omit&\cr
\tablerule
\omit&height2pt&\omit&&\omit&&\omit&\cr
&& 3 && $\frac{4 (b -1)}{(D-2) (b-2)}$ && $\eta_{\mu \nu}
\frac{\partial_{\rho} \partial_{\sigma}}{\partial^2} \!+\! \frac{\partial_{\mu}
\partial_{\nu}}{\partial^2} \eta_{\rho \sigma}$ & \cr
\omit&height2pt&\omit&&\omit&&\omit&\cr
\tablerule
\omit&height2pt&\omit&&\omit&&\omit&\cr
&& 4 && $a \!-\! 1$ && $4 \frac{\partial_{(\mu} \eta_{\nu) (\rho}
\partial_{\sigma)}}{\partial^2}$ & \cr
\omit&height2pt&\omit&&\omit&&\omit&\cr
\tablerule
\omit&height2pt&\omit&&\omit&&\omit&\cr
&& 5 && $-\frac{4 a (b-1)(b-3)}{(b-2)^2} \!-\! \frac{8 (b-1)(b + D - 3)}{
(b-2)^2 (D - 2)}$ && $\frac{\partial_{\mu} \partial_{\nu} \partial_{\rho}
\partial_{\sigma}}{\partial^4}$ & \cr
\omit&height2pt&\omit&&\omit&&\omit&\cr
\tablerule }}

\caption{Coefficient functions $\mathcal{C}_I(D,a,b)$ and the tensor
differential operators $[\mbox{}_{\mu\nu} \mathcal{T}^I_{\rho
\sigma}]$ for the graviton propagator (\ref{newGRprop}) defined with
the general gauge fixing functional (\ref{GRnew}).}

\label{CTs}

\end{table}

The most general Poincar\'e invariant extension of the graviton gauge 
fixing functional (\ref{GRfix}) depends upon two parameters $a$ and $b$,
\begin{equation}
\mathcal{L}_{\rm GRnew} = -\frac1{2a} \eta^{\mu\nu} \mathcal{F}_{\mu}
\mathcal{F}_{\nu} \qquad , \qquad \mathcal{F}_{\mu} \equiv \eta^{\rho\sigma}
\Bigl( h_{\mu \rho , \sigma} \!-\! \frac{b}{2} h_{\rho\sigma , \mu}\Bigr) 
\; . \label{GRnew}
\end{equation}
The associated propagator is \cite{DMC},
\begin{equation}
i\Bigl[\mbox{}_{\alpha\beta} \Delta^{\rm new}_{\gamma\delta}\Bigr](x;x') 
= \sum_{I=1}^5 \mathcal{C}_I(D,a,b) \times \Bigl[\mbox{}_{\mu\nu} 
\mathcal{T}^I_{\rho\sigma} \Bigr] \times i\Delta(x;x') \; ,
\label{newGRprop}
\end{equation}
where the coefficient functions $\mathcal{C}_I(D,a,b)$ and the tensor 
differential operators $[\mbox{}_{\mu\nu} \mathcal{T}^I_{\rho\sigma}]$ 
are given in Table~\ref{CTs}. The propagator can be given a more
revealing expression using the transverse projection operator
$\Pi_{\mu\nu} \equiv \eta_{\mu\nu} - \frac{\partial_{\mu} \partial_{\nu}}{
\partial^2}$,
\begin{eqnarray}
\lefteqn{i\Bigl[\mbox{}_{\alpha\beta} \Delta^{\rm new}_{\gamma\delta}
\Bigr](x;x') = \Biggl\{ 2 \Pi_{\mu (\rho} \Pi_{\sigma) \nu} \!-\! 
\frac2{D \!-\! 1} \Pi_{\mu\nu} \Pi_{\rho\sigma} } \nonumber \\
& & \hspace{.5cm} -\frac2{(D \!-\! 2)(D \!-\! 1)} \Biggl[ \eta_{\mu\nu}
\!-\! \Bigl( \frac{D b \!-\! 2}{b \!-\! 2} \Bigr) \frac{\partial_{\mu}
\partial_{\nu} }{\partial^2} \Biggr] \Biggl[ \eta_{\rho\sigma}
\!-\! \Bigl( \frac{D b \!-\! 2}{b \!-\! 2} \Bigr) \frac{\partial_{\rho}
\partial_{\sigma} }{\partial^2} \Biggr] \nonumber \\
& & \hspace{4.4cm} + 4a \times \frac{\partial_{(\mu} \Pi_{\nu) (\rho}
\partial_{\sigma)} }{\partial^2} + \frac{4 a}{(b \!-\! 2)^2} \times
\frac{\partial_{\mu} \partial_{\nu} \partial_{\rho} \partial_{\sigma}}{
\partial^4} \Biggr\} . \qquad
\end{eqnarray}
Of course the tranverse-traceless term on the first line represents the
contribution from dynamical, spin two gravitons. This term looms large
in the quantum gravity literature but it is well to recall that it plays
no role in the solar system tests of general relativity. The 
phenomenologically more important parts of the graviton propagator
are those on the second and third lines, which mediate the gravitational
interaction between sources of stress-energy. Note that the longitudinal 
terms proportional to the gauge parameter $a$ would vanish in the exact 
gauge $h^{\nu}_{~ \mu , \nu} = \frac{b}{2} h_{, \mu}$.

The most general Poincar\'e invariant extension of the photon gauge
fixing functional (\ref{EMfix}) depends upon a single parameter $c$,
\begin{equation}
\mathcal{L}_{\rm EMnew} = -\frac1{2c} (\partial^{\mu} A_{\mu})^2 \; .
\label{EMnew}
\end{equation}
The associated propagator is,
\begin{equation}
i\Bigl[ \mbox{}_{\rho} \Delta^{\rm new}_{\sigma} \Bigr](x;x') =
\Biggl[ \eta_{\rho\sigma} + (c \!-\! 1) \frac{\partial_{\rho} 
\partial_{\sigma}}{\partial^2} \Biggr] i\Delta(x;x') \; .
\end{equation}
The longitudinal term proportional to $c-1$ can make no contribution
to the general gauge vacuum polarization (\ref{ABdef}) because the
vertex function (\ref{Vvert}) is antisymmetric under interchange of 
its second and fourth indices,
\begin{equation}
V^{\mu\rho\kappa\lambda\alpha\beta} = -
V^{\mu\lambda\kappa\rho\alpha\beta} \; .
\end{equation}

It remains to explain how to act the tensor differential operators
of Table~\ref{CTs} on the scalar propagator (\ref{scalprop}). First 
note that inverse d'Alembertians act on $1/\Delta x^{D-2}$ to give,
\begin{eqnarray}
\frac1{\partial^2} \frac1{\Delta x^{D-2}} & = & -\frac1{2 (D\!-\! 4)} 
\frac1{\Delta x^{D -4}} \; , \\
\frac1{\partial^4} \frac1{\Delta x^{D-2}} & = & \frac1{8 (D\!-\! 4) 
(D \!-\! 6)} \frac1{\Delta x^{D -6}} \; .
\end{eqnarray}
Now just act the derivatives in the numerator to conclude,
\begin{eqnarray}
\frac{\partial_{\mu} \partial_{\nu}}{\partial^2} \, i\Delta(x;x') & = &
\frac12 \times \Biggl\{ \eta_{\mu\nu} \!-\! \frac{ (D \!-\! 2) \Delta x_{\mu}
\Delta x_{\nu} }{\Delta x^2} \Biggr\} \, i\Delta(x;x') \; , \\
\frac{\partial_{\mu} \partial_{\nu} \partial_{\rho} \partial_{\sigma}}{
\partial^4} \, i\Delta(x;x') & = & \frac18 \times \Biggl\{ 3 
\eta_{(\mu\nu} \eta_{\rho\sigma)} \!-\! \frac{ 6 (D \!-\! 2) \eta_{(\mu\nu}
\Delta x_{\rho} \Delta x_{\sigma)}}{\Delta x^2} \\
& & \hspace{1cm} + \frac{D (D\!-\!2) \Delta x_{\mu} \Delta x_{\nu} 
\Delta x_{\rho} \Delta x_{\sigma}}{\Delta x^4} \Biggr\} \, i\Delta(x;x') 
\; . \qquad
\end{eqnarray}

\subsection{Gauge Dependent Proportionality Constant}

\begin{table}

\vbox{\tabskip=0pt \offinterlineskip
\def\tablerule{\noalign{\hrule}}
\halign to390pt {\strut#& \vrule#\tabskip=1em plus2em&
\hfil#\hfil& \vrule#& \hfil#\hfil& \vrule#&
\hfil#\hfil& \vrule#& \hfil#\hfil& \vrule#\tabskip=0pt\cr
\tablerule
\omit&height4pt&\omit&&\omit&&\omit&&\omit&\cr
\omit&height2pt&\omit&&\omit&&\omit&&\omit&\cr
&& $\!\!\!\! I \!\!\!\!$ && $\!\!\!\! A_I \!\!\!\!$
&& $\!\!\!\! B_I \!\!\!\!$ && $\!\!\!\! C_I \!\!\!\!$ & \cr
\omit&height4pt&\omit&&\omit&&\omit&&\omit&\cr
\tablerule
\omit&height2pt&\omit&&\omit&&\omit&&\omit&\cr
&& 1 && $\frac12 D (3 D \!-\! 8)$ && $3 D \!-\! 8$ &&
$\frac12 (D \!-\! 2)^2 (3 D \!-\! 8)$ & \cr
\omit&height2pt&\omit&&\omit&&\omit&&\omit&\cr
\tablerule
\omit&height2pt&\omit&&\omit&&\omit&&\omit&\cr
&& 2 && $\frac14 (D \!-\! 4)^2 D$ && $\frac12 (D \!-\! 4)^2$ &&
$\frac14 (D \!-\! 4)^2 (D \!-\! 2)^2$ & \cr
\omit&height2pt&\omit&&\omit&&\omit&&\omit&\cr
\tablerule
\omit&height2pt&\omit&&\omit&&\omit&&\omit&\cr
&& 3 && $\!\!\!\! \frac12 (D \!-\! 4)^2 (D \!-\! 1) \!\!\!\!$ && $D \!-\! 4$ &&
$\!\!\!\!\frac12 (D \!-\! 4) (D \!-\! 2)^2 (D \!-\! 1) \!\!\!\!$ & \cr
\omit&height2pt&\omit&&\omit&&\omit&&\omit&\cr
\tablerule
\omit&height2pt&\omit&&\omit&&\omit&&\omit&\cr
&& 4 && $(D \!-\! 1) (3 D \!-\! 8)$ && $D^2 \!-\! 2 D \!-\! 2$ &&
$(D \!-\! 2)^2 (D \!-\! 1)$ & \cr
\omit&height2pt&\omit&&\omit&&\omit&&\omit&\cr
\tablerule
\omit&height2pt&\omit&&\omit&&\omit&&\omit&\cr
&& 5 && $\!\!\!\! \frac18 (D \!-\! 2) (D \!-\! 1) D \!\!\!\!$
&& $\!\!\!\! \frac14 (D\!-\!2)(D\!-\!1) \!\!\!\!$
&& $\frac18 (D \!-\! 2)^3 (D \!-\! 1)$ & \cr
\omit&height2pt&\omit&&\omit&&\omit&&\omit&\cr
\tablerule}}

\caption{Coefficients $A_I$, $B_I$ and $C_I$ defined in relations
(\ref{ABdef}) and (\ref{Cdef}), under the replacement
$i[\mbox{}_{\mu\nu} \Delta^{\rm new}_{\rho\sigma}](x;x') \longrightarrow
[\mbox{}_{\mu\nu} \mathcal{T}^I_{\rho\sigma}] \times i\Delta(x;x')$ for 
each of the five tensor differential operators defined in Table~\ref{CTs}.}

\label{ABC}

\end{table}

We are now ready to compute the crucial proportionality constant 
of relation (\ref{genform}). Because the gauge dependence of the
photon propagator drops out, we need only consider the gauge
dependence of the graviton propagator. Because the graviton propagator
(\ref{newGRprop}) is a sum of gauge-dependent coefficients 
$\mathcal{C}_I(D,a,b)$ times tensor operators $[\mbox{}_{\mu\nu}
\mathcal{T}^I_{\rho\sigma}]$, acting on the scalar propagator, we
may as well work out the result for each tensor operator separately.
Table~\ref{ABC} presents the coefficients $A_I(D)$ and $B_I(D)$ which were 
defined in relation (\ref{ABdef}), for each of the five tensor differential
operators of Table~\ref{CTs}. Also given is the contribution of each
tensor differential operator to the coefficient $C_I(D)$,
\begin{equation}
C_I(D) \equiv D \times A_I(D) - 2 (D \!-\! 1) \times B_I(D) \; .
\end{equation}

To recover the full result for $C(D,a,b)$ defined in relation (\ref{Cdef})
we multiply each $C_I(D)$ by the appropriate gauge dependent
coefficient $\mathcal{C}(D,a,b)$ from Table~\ref{CTs},
\begin{equation}
C(D,a,b) = \sum_{I=1}^5 \mathcal{C}_I(D,a,b) \times C_I(D)
\end{equation}
The formula for arbitrary $D$ is not illuminating, but specializing to
$D = 4$ gives,
\begin{equation}
C(4,a,b) = -12 a \times \frac{(3 b^2 \!-\! 12 b \!+\! 8)}{(b \!-\! 2)^2}
- \frac{4}{(b \!-\! 2)^2} \; .
\end{equation}
Our original gauge corresponds to $a=b=1$, which gives $C(4,a,1) = 8$.
Hence the various one loop corrections computed in section 3 are valid 
for general $a$ and $b$ if we multiply by the proportionality constant,
\begin{equation}
K(a,b) \equiv \frac{C(4,a,b)}{C(4,1,1)} = -\frac{3}2 a \times
\frac{(3 b^2 \!-\! 12 b \!+\! 8)}{(b \!-\! 2)^2} - 
\frac1{2 (b \!-\! 2)^2} \; . \label{Kab}
\end{equation}
It is interesting to note that the gauge independent contribution from
dynamical gravitons vanishes in $D=4$ dimensions,
\begin{eqnarray}
\lefteqn{C_1(D) \!-\! C_4(D) \!+\! 2 C_5(D) \!-\! \frac2{D \!-\! 1} 
\Bigl[C_2(D) \!-\! C_3(D) \!+\! C_5(D)\Bigr] } \nonumber \\
& & \hspace{6cm} = \frac{(D \!-\! 4) (D\!-\! 2)^2 (D \!+\! 1) 
(D \!+\! 2)}{4 (D \!-\! 1)} \; . \qquad
\end{eqnarray} 

It is apparent that the gauge dependent proportionality constant 
(\ref{Kab}) can be made to have either sign by varying the gauge
parameter $a$. Furthermore, $K(a,b)$ can be made arbitrarily large 
in magnitude by taking the gauge parameter $b$ close to $2$. Hence 
it would seem that our results are completely gauge dependent and
unphysical. Gauge dependence has also been noted in the renormalization
group approach \cite{ARP,DJT}.

A moment's thought reveals that all is not lost because {\it the 
result of Bjerrum-Bohr} \cite{Coulomb} {\it for the quantum gravitational 
correction to the Coulomb potential was derived from the gauge 
independent S-matrix of scalar QED.} Roughly speaking, this correction
derives from the fact that gravity is sourced by the electromagnetic 
fields of the two charged particles being scattered, and this source 
changes as the particles move with respect to one another. That is a real 
effect, not some gauge artifact. And it is crucially important to note
that {\it we agree with Bjerrum-Bohr up to a factor of} $+9$. We 
attributed this factor to our having only quantum-corrected the left 
hand side of the operator Maxwell equation,
\begin{equation}
\partial^{\nu} \Bigl[ \sqrt{-g} \, g^{\nu\rho} g^{\mu\sigma} 
F_{\rho\sigma}(x) \Bigr] = J^{\mu}(x) \; .
\end{equation}
The right hand side is also an operator and it must also suffer
quantum gravitational corrections, one of which is depicted in 
Fig.~\ref{vertex}. By Poincar\'e invariance, current conservation and 
dimensional analysis, those corrections must take exactly the same 
form as we found for the left hand side, up to an overall constant.
We conjecture that the gauge dependence $K(a,b)$ we have just found for 
corrections to the left hand side is canceled by gauge dependence in
corrections to the right hand side. If this is correct, then {\it the 
overall, gauge independent correction to the various results derived in 
section 3 can be inferred by comparing any one of them with its S-matrix 
analogue}. For scalar QED the correction factor would be 9, and it could
be computed for the point particle source we used.

The resolution we have just proposed to the gauge issue recalls some old 
work by DeWitt \cite{BSDW1} about dependence upon the gauge fixing 
functionals even in the gauge invariant background field effective 
action $\Gamma = S + \Sigma$. DeWitt states \cite{BSDW2}, ``The functional 
form of $\Sigma$ is not independent of the choice of these terms. However, 
the solutions of the {\it effective field equation} can be shown to be the 
same for all choices.'' At the order we are working there is no distinction
between our effective field equations and those of the gauge invariant
background field effective action. (One can see this from the transversality
of our vacuum polarization.) However, we have just shown that DeWitt's
statement cannot be correct if the source (or the asymptotic field
strengths for scattering solutions) is not normalized in some physical way. 
Our proposal is that including quantum corrections to the right hand side 
of the equation provides this physical normalization. More work is
obviously required, in particular an explicit computation of the quantum
gravitational corrections to the source, but it would be wonderful if
solutions to the effective field equations could be physically interpreted 
the same way as classical solutions.

\section{Discussion}

We used dimensional regularization to compute the one loop quantum
gravitational contribution (\ref{transverse}) to the vacuum
polarization on flat space background. A fully renormalized result
(\ref{renorm}) was obtained by first partially integrating to
localize the ultraviolet divergence and then absorbing it into the
appropriate BPHZ counterterm (\ref{ctm}) with coefficient
(\ref{C4}). The Schwinger-Keldysh formalism \cite{earlySK,lateSK}
was then employed to reach the manifestly real and causal form
(\ref{finalform}).

We used (\ref{finalform}) to solve the quantum corrected Maxwell's
equation (\ref{qmax1}) for various special cases. Provided the
appropriate perturbative corrections to the initial state cancel the
surface terms involved in reaching the form (\ref{qmax2}), there is
no change in the source-free solutions at any order in the loop
expansion. However, sources induce a variety of interesting effects.

Probably the most provocative source is the current density
(\ref{J4Idipole}) of an instantaneously created, point electric
dipole. The pulse (\ref{smeared}) which results in the magnetic
field propagates slightly outside the classical light-cone. It seems
to arise from quantum fluctuations of the metric operator, which are
isotropic but favor super-luminal propagation because there is more
volume outside the light-cone than inside. That this sort of thing
might occur has been realized since the earliest days of quantum
gravity \cite{smear}. Our super-luminal effect is completely Lorentz
invariant, merely changing the characteristic surface from $\eta_{\mu\nu}
x^{\mu} x^{\nu} = 0$ to $\eta_{\mu\nu} x^{\mu} x^{\nu} = 
\frac{4G}{3\pi}$. Despite many claims to the contrary, this seems to 
be the first case of super-luminal propagation from a quantum field 
theory whose classical analogue does not allow super-luminal propagation.
All previous claims have been based on derivative expansions \cite{DH,DS}, 
which are perfectly valid for most applications of low energy effective 
field theory but which incorrectly treat the high frequency modes needed 
to resolve the propagation of a pulse.

The other interesting source we studied is the response to a static,
point charge. Our result (\ref{Cpot}) for the quantum corrected Coulomb 
potential agrees with what Radkowski found more than four decades ago
\cite{AFR}. It does not agree with the Coulomb potential Bjerrum-Bohr 
inferred from the scattering of charged, gravitating scalars 
\cite{Coulomb}, but we all find quantum gravity strengthens the
electrostatic force at short distances. We believe that the factor of 
nine discrepancy with Bjerrum-Bohr derives from his S-matrix technique
implicitly including quantum gravitational corrections to the charge
density, like the diagram depicted in Fig.~\ref{vertex}. This should 
have been done for our point particle source, but it would only have
changed the one loop field strengths by an overall constant.

Renormalization group analyzes \cite{RW,Reuter1,Reuter2} seem to
provide a more serious discrepancy. These studies find that quantum 
gravity reduces the electrodynamic coupling constant at large scales. 
The usual inference would be that quantum gravity weakens the 
electrostatic force at short distances. However, the beta function is
not itself observable; several other effects must be combined to infer
the impact of quantum gravity. The S-matrix computation of Bjerrum-Bohr 
should include all of these effects, and it shows that quantum gravity 
strengthens the electrostatic force at short distances.

Gauge dependence poses a major obstacle to the physical interpretation
of solutions to the effective field equations. If one restricts to 
Poincar\'e invariant gauge fixing functionals, the only possible change
to our vacuum polarization (\ref{finalform}) is rescaling by an overall,
gauge-dependent constant. In section 4 we considered the most general
2-parameter family of graviton gauges (\ref{GRnew}), and the most general
1-parameter family of photon gauges (\ref{EMnew}). We showed that the
vacuum polarization has no dependence upon the electromagnetic gauge
fixing parameter $c$, but it depends strongly on the two gravitational
gauge fixing parameters $a$ and $b$. The effect of being in a general
covariant gauge is to rescale (\ref{finalform}) by the function $K(a,b)$
given in equation (\ref{Kab}). By varying the constants $a$ and $b$,
one can make $K(a,b)$ assume any values from plus infinity to minus 
infinity.

Such massive gauge dependence would seem to invalidate any physical
inference from the results of section 3, however, the gauge independent
result of Bjerrum-Bohr suggests a simple resolution. There is no question
that one must include quantum gravitational corrections to the current
density operator. This seems to be why Bjerrum-Bohr (who implicitly did
this) gets a factor of nine different one loop correction to the Coulomb
potential. We conjecture that making such corrections in a general gauge
--- which seems quite feasible using the techniques of section 4 ---
would completely cancel the gauge dependence of our result. If this
could be demonstrated then it would be possible to realize the old
dream \cite{BSDW1,BSDW2} of using solutions to the effective field 
equations as freely as one does classical solutions. {\it Note also that
it would provide an important class of observables in cosmology, for
which the S-matrix does not exist.}

The point of this exercise has been to establish the flat space
correspondence limit for a planned investigation of the effects of
inflationary gravitons on electromagnetism. Our model has been a
similar study of the effects of inflationary scalars on gravity
\cite{PW2,PW3}, the flat space limit of which \cite{PW1} played a
crucial role in guiding the analysis. In retrospect, we can
recognize the simplicity of flat space as the ideal venue for
sorting out the troublesome issues of dependence upon the choice
of field variable and the choice of gauge which are so important
to a correct interpretation of the many solutions which now exist
to linearized effective field equations on de Sitter background
\cite{SQED,Henri,BOW,TW3,Yukawa,Miao,Kahya1,Kahya2}.

\vskip .5cm

\centerline{\bf Acknowledgements}

We have profited from discussion and correspondence on this subject
with A. Barvinsky, S. Deser, L. H. Ford, G. 't Hooft, B. L. Hu, D.
Mazzitelli and M. Reuter. This work was partially supported by NSF
grant PHY-0855021 and by the Institute for Fundamental Theory at
the University of Florida.

\end{document}